\documentclass[lettersize,journal]{IEEEtran}
\usepackage{amsmath,amsfonts}
\usepackage{algorithmic}
\usepackage{algorithm}
\usepackage{array}
\usepackage[caption=false,font=normalsize,labelfont=sf,textfont=sf]{subfig}
\usepackage{textcomp}
\usepackage{stfloats}
\usepackage{url}
\usepackage{verbatim}
\usepackage{graphicx}
\usepackage{cite}
\usepackage{booktabs}
\usepackage{multirow}
\usepackage{color}
\usepackage{makecell}
\usepackage{array}
\usepackage[dvipsnames]{xcolor}

\usepackage{hyperref}

\begin{document}

\title{Efficient LLM-based Subgraph Retrieval for Multi-Hop Knowledge Base Question Answering}
\author{Duanyang Yuan, $\text{Sihang Zhou}^*$, Xiaoshu Chen, Haoyuan Chen, Ke Liang, $\text{Jian Huang}^*$
\thanks{Duanyang Yuan, Sihang Zhou, Haoyuan Chen, and Jian Huang are with the College of Intelligence Science and Technology, National University of Defense Technology, Changsha 410073, China, and also with the National Key Laboratory of Equipment State Sensing and Smart Support, Changsha 410073, China. (e-mail: \{shzhou, huang\_jian\}@nudt.edu.cn) \par
Xiaoshu Chen and Ke Liang are with the College of Computer Science and Technology, National University of Defense Technology, Changsha 410073, China  \par
$*$ Corresponding Authors}
}

\markboth{IEEE TRANSACTIONS ON KNOWLEDGE AND DATA ENGINEERING}%
{Shell \MakeLowercase{\textit{et al.}}: A Sample Article Using IEEEtran.cls for IEEE Journals}


\maketitle

\begin{abstract}
Multi-hop Knowledge Base Question Answering (KBQA) aims to find answer entities in the knowledge base that are multiple hops away from the entities in the question. Information retrieval-based (IR-based) methods extract a pivotal subgraph from the entire KB to locate candidate answers and then evaluate their plausibility through semantic matching with the question. However, we observed that the extracted subgraphs often include nodes that are weakly related or irrelevant to the question. Without a proper node filtering mechanism, the number of irrelevant nodes grows as the number of hops increases, leading to excessive consumption of computational resources.
To address these challenges, this study introduces an efficient LLM-based subgraph retrieval method for multi-hop knowledge base question answering, M-ER. The framework leverages Monte Carlo Tree Search (MCTS) to transform subgraph exploration into a tree-structured search process. During the MCTS selection phase, nodes that are highly relevant to the question are prioritized for inclusion in the subgraph, eliminating the need to traverse all nodes in the KB. The framework further incorporates a large language model (LLM) to refine the search direction, ensuring that exploration remains focused on nodes relevant to the question. In addition, selected nodes are quantitatively scored, and these scores are fed back into the node selection process to effectively filter out irrelevant candidates, thereby improving the quality of the subgraph. This mechanism not only narrows the search space but also enhances the overall efficiency of multi-hop KBQA.
Experiments on the WebQSP benchmark demonstrate that M-ER achieves 78.88\% on the Hits@1 metric, while also improving computational efficiency. These results not only validate the effectiveness of M-ER, but also offer a viable technical path to balance performance and computational efficiency\footnote{https://github.com/DuanyangYuan/M-ER}.
\end{abstract}

\begin{IEEEkeywords}
Multi-hop Question Answering, MCTS, Knowledge Base, Large Language Model.
\end{IEEEkeywords}

\section{Introduction}
\IEEEPARstart{k}{nowledge} base question answering (KBQA) \cite{1} aims to retrieve answers from structured knowledge bases (e.g., Freebase \cite{2}, DBpedia \cite{3}) in response to natural language queries. Current methodologies can be divided into two paradigms: semantic parsing-based (SP-based) \cite{4, 5, 45, 77} and information retrieval-based (IR-based) \cite{8,9}. SP-based approaches translate questions into executable logical forms \cite{45, 91} but require expensive annotation efforts. In contrast, IR-based methods directly represent and rank candidate entities based on their relevance to the input questions. 

\begin{figure*}[t] 
    \centering
    \includegraphics[scale=0.6]{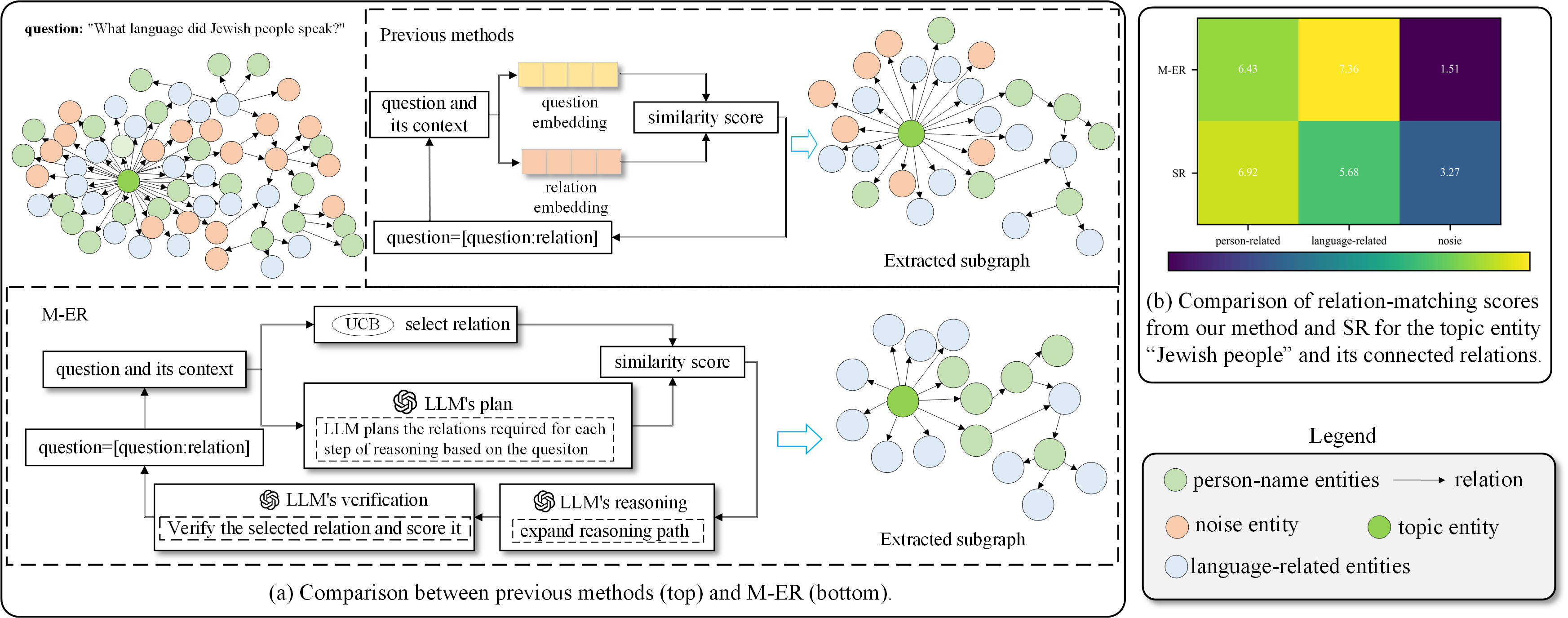} 
    \caption{Illustrates the challenges faced by previous methods (SR) and how M-ER alleviates them. (a) A comparison between SR (top) and M-ER (bottom) is presented. SR expands subgraphs primarily based on semantic similarity, which tends to introduce nodes that are weakly related or irrelevant to the question. These irrelevant expansions distract the reasoning process and incur unnecessary computational overhead. In contrast, M-ER incorporates LLM-based reflection, reasoning, and evaluation stages. M-ER evaluates each candidate node for expansion by considering both the current subgraph context and the question, thereby explicitly estimating its potential contribution to answering the question. (b) Comparison of relation-matching scores from our method (M-ER) and SR. Specifically, M-ER computes similarity between the predicted relation and each relation linked to the topic entity, and SR measures similarity between the query and those relations. While SR struggles to distinguish answer-relevant relations from those that are weakly related or irrelevant to the question, M-ER effectively prioritizes relations that are answer-relevant (e.g., \textit{language-related}) and suppresses weakly relevant relations (e.g., \textit{person-related}).}
    
    \label{intro1}
\end{figure*}

\begin{table*}[]
    \centering
    \caption{The one-hop neighbors of the topic entity \textit{Jewish people} are selected by calculating similarity scores between the question and the relations. Only those neighbors that meet the predefined threshold can be added to the subgraph. The term “Frequency” indicates how often the one-hop relations appears in the dataset.}
    \begin{tabular}{cccc}
     \hline
    \multicolumn{2}{l}{
    \begin{tabular}[c]{@{}l@{}}
        Question:   What language did Jewish people speak?\\ 
        Topic entity: Jewish people  \\ 
        The number of one-hop neighbor relations of the topic entity: 5308  
    \end{tabular}} \\ \hline
    Relation Name          & Frequency                                                    &    Relation Name      & Frequency    \\ \hline
    people.ethnicity.people  & 5149                                                      &      people.ethnicity.includes\_groups   & 13     \\
    common.topic.webpage        & 47                                                    &  book.book\_subject.works         &12  \\
    people.ethnicity.languages\_spoken       & 25                                       & military.military\_combatant.military\_conflicts  & 4 \\
    fictional\_universe.ethnicity\_in\_fiction.characters\_of\_this\_ethnicity     & 25      & film.film\_subject.films    & 3    \\
    type.object.type  & 16                                                             & common.topic.image & 2                         \\  \hline
    \end{tabular}
    \label{intro}
\end{table*}
Extracting a question-relevant subgraph and reasoning within it is a key branch of IR-based methods. The performance of subgraph extraction is mainly influenced by two factors: subgraph construction quality and answer coverage rate. Subgraph construction directly affects the efficiency and accuracy of the reasoning process, as reasoning paths are formed by expanding nodes and relations within the subgraph. The answer coverage rate defines the upper bound of the performance \cite{10}. 
Current subgraph extraction methods can be categorized into two main approaches: the k-hop neighbor method and the scoring-based method. The k-hop neighbor approach constructs subgraphs by retrieving a fixed number of hops around the topic entity without requiring additional search heuristics. However, KBQA is sensitive to subgraph size, since a small subgraph may exclude correct answers, while a large one may introduce excessive noise. Scoring-based methods perform KBQA by quantifying the semantic association between the question and KB elements. NSM \cite{6} adopts a teacher-student framework with bidirectional reasoning to generate intermediate supervision signals for answer extraction. SR \cite{44} introduces a trainable subgraph retriever decoupled from inference, and generates accurate question-related subgraphs through weakly supervised pre-training and end-to-end fine-tuning. When combined with NSM \cite{6}, this approach sets a new performance benchmark.

Existing scoring-based methods typically construct subgraphs by computing semantic similarity between the question and knowledge base elements (entities or relations), as shown in Figure~\ref{intro1}(a). Representative methods such as SR \cite{44} use language models to compute semantic similarity. However, when a single entity has relations corresponding to diverse attributes, these methods may retain relations that are weakly related or irrelevant to the question, expanding the search space and reducing answer accuracy. For example, for the question \textit{What language did Jewish people speak?}, we analyze the total number of one-hop neighboring relations of the topic entity \textit{Jewish people}, along with the names and frequencies of the neighboring relations filtered based on similarity, as shown in Table~\ref{intro}. From this, three issues can be observed: (1) The topic entity \textit{Jewish people} has rich attribute information, linking to language-related entities via the relation \textit{people.ethnicity.languages\_spoken} and to person-related entities via the relation \textit{people.ethnicity.people}. All these links originate from a single entity. (2) Existing methods typically compute similarity between the query [\textit{What language did Jewish people speak?; Jewish people}] and all relations connected to the topic entity \textit{Jewish people}. As a result, the person-related relation \textit{people.ethnicity.people} may be retained during retrieval even though only language-related information is relevant to answering the question. Notably, person-related relations (e.g., \textit{people.ethnicity.people}) account for 97\%, far exceeding relevant language-related relations (e.g., \textit{people.ethnicity.languages\_spoken}), thereby overshadowing crucial language-related links and propagating semantic noise to subsequent hops. This propagation can lead to a combinatorial explosion, reducing overall performance. (3) The topic entity \textit{Jewish people} has 5,308 one-hop neighboring relations. Extending the subgraph based on this introduces substantial computational overhead. 

To address these challenges, Coke \cite{81} formulates reasoning as a trade-off between accuracy and computational cost, aiming to achieve cost-effective inference. Although effective in principle, this approach relies on global answer matching over the entire KB, which introduces substantial noise and increases the overall computational burden. MCTS provides a more targeted exploration strategy \cite{22, 23, 17}. By simulating future states through random sampling, it incrementally constructs a search tree where the nodes correspond to possible reasoning steps within the subgraph. This strategy focuses computation on the most promising nodes, reducing the need for exhaustive enumeration. By balancing exploration and exploitation, MCTS narrows the scope of the search and improves computational efficiency \cite{14, 15, 16, 17}. Extending this idea, KBQA-o1 \cite{86} leverages MCTS to balance exploration quality and search range, while also adopting incremental fine-tuning to reduce reliance on labeled data. Its limitation lies in disregarding the structural and semantic context of the knowledge graph, making it vulnerable to cascading errors from incorrect intermediate predictions. MCTS-KBQA \cite{88} introduces a task-specific scoring function to improve answer selection. However, it lacks explicit guidance for selecting subsequent search directions and does not fully exploit the structural information within subgraphs. Consequently, it often terminates at intermediate states and treats them as final answers, leading to lower accuracy and unstable performance. In addition, MCTS often struggles in scenarios where a single entity is connected to relations corresponding to diverse attributes, as weakly related or irrelevant nodes may be retained during exploration. This reduces the model’s ability to discriminate answer-relevant semantics, leading to the inclusion of noisy entities and a degradation in overall performance. Therefore, M-ER is designed to incorporate MCTS into KBQA while explicitly filtering irrelevant relations.

Large language models (LLMs), on the other hand, exhibit strong contextual reasoning capabilities \cite{109, 110} and can guide node selection toward answer-relevant directions \cite{14, 18}. Therefore, this study proposes an efficient LLM-based subgraph retrieval method for multi-hop knowledge base question answering (M-ER), which integrates the semantic reasoning capabilities of LLMs with the global search efficiency of MCTS, as shown in Figure \ref{intro1}(a). This approach jointly activates model capabilities from two perspectives: internal (semantic reasoning within the model) and external (exploration of the KB). Specifically, LLMs interpret the question to predict the next search direction, which is then matched with candidate nodes based on semantic similarity. Meanwhile, selected nodes are evaluated based on their contribution to the final answer, and these scores are used to refine the reflection process, thus improving the alignment between the model’s internal knowledge and external information. It can be seen from Figure \ref{intro1} (b) that this method can effectively reduce the similarity scores of irrelevant nodes, thus improving the quality of the subgraph. The main contributions of this work are in three aspects:

\begin{itemize}
  \item We combine MCTS and LLM for subgraph construction and solve the trade-off between semantic focus and search efficiency by balancing exploration and exploitation. Experimental results on WebQSP demonstrate that M-ER achieves 78.88\% on the Hits@1 metric, surpassing the previous relevance-driven method (SR) by 7.08\%.
  \item We design a guidance mechanism where LLMs steer MCTS toward answer-relevant subgraph expansions, effectively mitigating semantic drift.
  \item We develop a subgraph evaluation mechanism that assigns quantitative scores to selected nodes, enabling feedback-driven optimization of reasoning paths within the subgraph. This mechanism opens new avenues for KBQA.
\end{itemize}

\section{Related Work}
\subsection{Subgraph Retrieval-based KBQA}
We review subgraph retrieval methods in KBQA from two aspects: the k-hop neighbor method and the scoring-based method. 
\subsubsection{K-hop neighbor methods} 
The k-hop neighbor method can extract subgraphs in two ways. The k-hop neighbors of the topic entity are selected directly as a subgraph. The other method (e.g., RASR \cite{10}) first computes similarity between the question and all relations. It then selects the most relevant relations to construct the initial 1-hop subgraph, which is further expanded to form the 2-hop subgraph. However, this rigid hop-based selection may exclude correct answers when the hop size is too small, while introducing substantial noise when the hop size is too large.

\subsubsection{Scoring-based methods} 
The methods perform KBQA by quantifying the semantic association between the question and the KB elements. The key is to design a similarity scoring function to screen the key information. The early key-value memory network \cite{40} stores context as key–value pairs, enabling the model to learn how keys can be used to retrieve values relevant to the question. However, this method is limited by hardware memory. With the development of end-to-end neural networks, GRAFT-Net \cite{24} retrieves the nodes and relations related to the question in the KB to construct a subgraph containing high-probability answers. Node representations are then learned over the constructed subgraph, based on which answer candidates are inferred. This method lacks supervision and thus cannot constrain the reasoning process. To address this limitation, some studies propose improvements, such as PullNet \cite{5}, SR \cite{44}, FC-KBQA \cite{59}, CBR-SUBG \cite{60}, and Query Graph Ranking \cite{76}. However, these methods primarily focus on the similarity between the question and the KB, without considering whether the retrieved nodes actually contribute to answering the questions. As a result, irrelevant information that is similar to the question is introduced, leading to subgraph redundancy and unnecessary consumption of computational resources.

\subsection{LLM-based KBQA}
The advent of LLMs has greatly improved the reliability of KBQA performance \cite{107, 108}. The LLM-based KBQA methods can generally be categorized into several types. The first category adopts the generation-retrieval paradigm, in which LLM generates the initial answer or logical rules, and then refines it using entity/relation retrieval or document verification (e.g., ChatKBQA \cite{64}, FuSIC-KBQA \cite{65}, GenGround \cite{63}, RoG \cite{90}). ChatKBQA \cite{64} employs an innovative generate-then-retrieve approach. A fine-tuned LLM first produces a skeleton of a logical form, which is then aligned with the knowledge graph via unsupervised retrieval of entities and relations. RoG \cite{90}, on the other hand, integrates structural knowledge from KGs through its planning–retrieval–reasoning framework, generating KG-grounded relation paths as faithful plans for interpretable reasoning. Although these methods reduce dependence on explicit semantic parsers, they remain sensitive to retrieval noise and incur high computational costs.

The second category focuses on question decomposition and step-by-step reasoning (e.g. ICSU \cite{78}, GRV-KBQA \cite{95}), thereby enhancing interpretability. GRV-KBQA \cite{95} adopts a three-stage framework that decouples logical structure generation, semantic grounding, and structure-aware verification. It first decomposes a complex query into a chain of subquery templates and then fills these templates to produce executable logical forms for question answering. However, such decomposition methods are limited when relevant information exceeds the predefined query scope. Other methods, such as RGR-KBQA \cite{94}, follow a “retrieve–generate–retrieve” three-stage framework. In this approach, cosine similarity is used to extract the most relevant relation chains, which enhances semantic understanding and improves generation accuracy.
Nonetheless, most of these methods are heavily dependent on similarity metrics and prompt engineering, which introduces substantial noise and hampers overall performance. 
Currently, LLM-based KBQA focuses mainly on how to make full use of the inherent knowledge and the strong reasoning ability of LLM in a large KB, while ignoring the role of subgraphs and efficient search strategies. Therefore, our work is dedicated to bridging these knowledge gaps.

\section{Method}

\begin{figure*}[t] 
    \centering
    \includegraphics[scale=0.60]{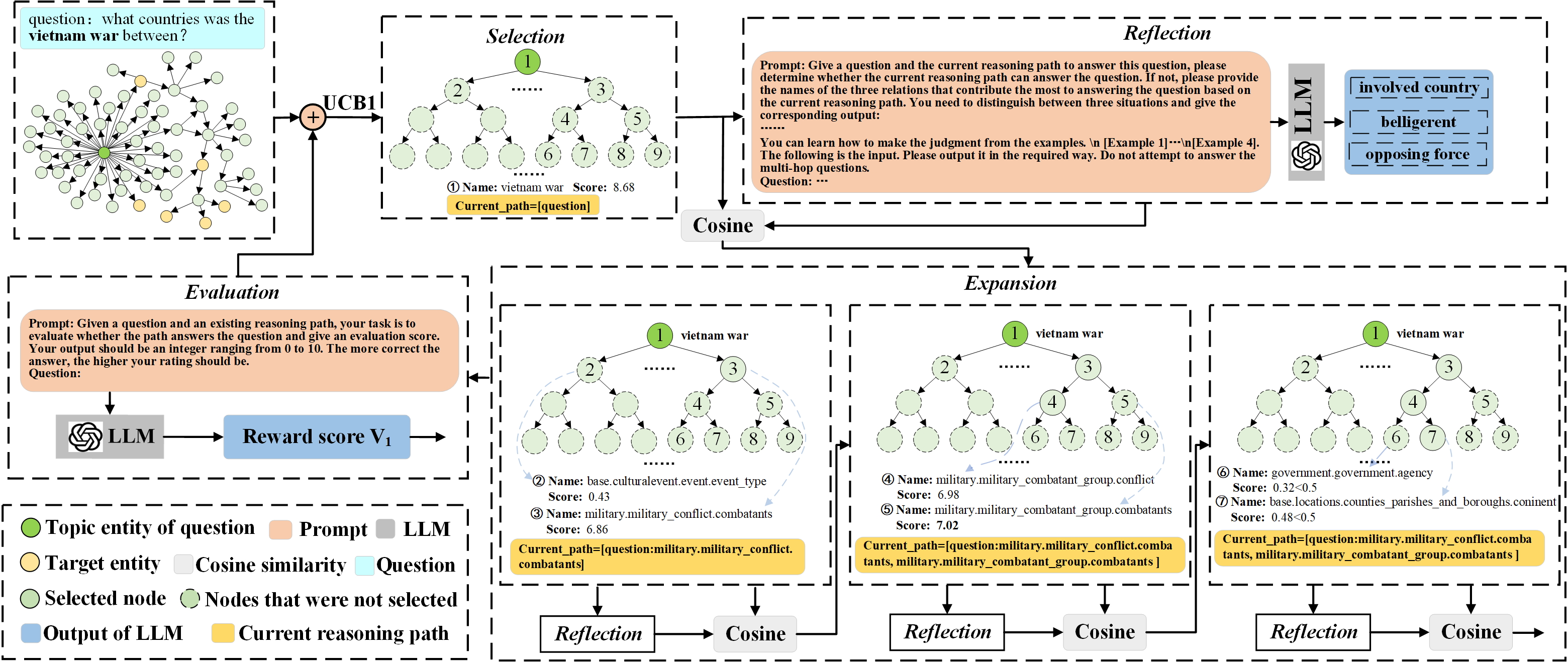} 
    \caption{The framework of M-ER. During the selection phase, the LLM-guided reward score optimizes the node exploration strategy, aiming to select the node most relevant to answering the question while minimizing the inclusion of irrelevant nodes. In the reflection stage, the LLM assesses the reasoning paths within the current subgraph and suggests several candidate nodes for the next search direction. The expansion phase leverages both the LLM-guided reflection and the existing KB to collaboratively determine the next search direction. This process enables the generation of multiple reasoning paths, which go beyond the constraints of manual rules. In the evaluation phase, the nodes identified in the expansion phase are scored. The process is simulated iteratively, with each cycle updating the reward score recursively to the UCB1 value of the parent node.}
    \label{method}
\end{figure*}

Extracting and reasoning over a subgraph that is relevant to the question is a typical approach in KBQA. To extract valuable information from the graph and minimize information loss, we propose an efficient LLM-based subgraph retrieval method for multi-hop knowledge base question answering, called M-ER. The overall framework is illustrated in Figure~\ref{method}.
In this process, the subgraph not only provides the information required to answer the question but also captures the reasoning paths connecting the topic entity to the answer node. Each iteration builds on this subgraph to produce a specific reasoning path. By exploring these expanding subgraphs through a tree structure, the LLM can introduce sufficient diversity, thereby fully activating its reasoning capabilities.

\subsection{Predefined Relations in Subgraph}

\begin{figure}
\centering
\includegraphics[scale=1]{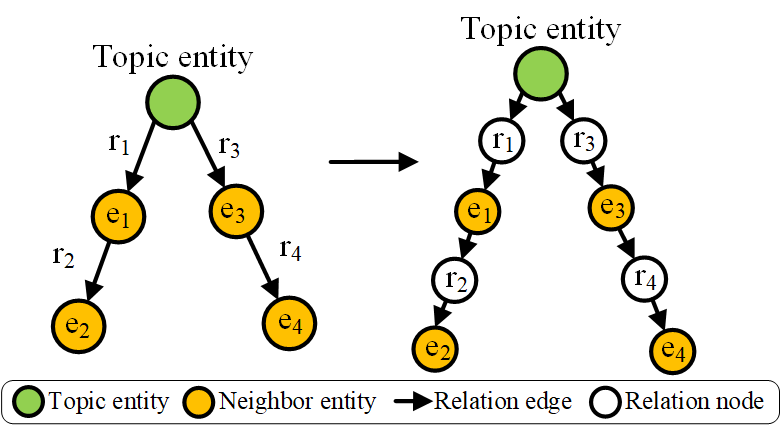}
\caption{Transforming relation edges into relation nodes. In the left subfigure, $e_1, e_2, e_3,$ and $e_4$ denote entity nodes. After transformation, as shown in the right subfigure, $r_1, r_2, r_3,$ and $r_4$ denote relation nodes. Specifically, we regard the topic entity node as the 0-hop node.}
\label{relaton_def}
\end{figure}

As shown in Figure \ref{relaton_def}, before the transformation, $e_1$, $e_2$, $e_3$, and $e_4$ denote neighboring entity nodes, while $r_1$, $r_2$, $r_3$, and $r_4$ denote relation edges connecting the topic entity to its neighbors. After the transformation, relation edges are reified as relation nodes. Consequently, $r_1$, $r_2$, $r_3$, and $r_4$ become explicit nodes in the subgraph. Transforming relation edges into relation nodes provides significant advantages for subgraph reasoning tasks~\cite{30}. By reifying relations as nodes, the graph explicitly encodes second-order semantic associations through direct topological connections between relation nodes. As illustrated in Table~\ref{intro}, when handling queries such as \textit{What language did Jewish people speak?}, the relation-node formulation enables direct similarity computation between the question semantics and relation nodes (e.g., \textit{people.ethnicity.languages\_spoken}). This design reduces the semantic deviation introduced by traditional methods that require indirect matching through entity intermediaries.


\subsection{Selection Based on Reward Score}

The process begins by selecting a node $Leaf_{i}$ from the tree hierarchy. In each node selection, the candidate nodes are evaluated using the Upper Confidence Bound (UCB1) metric \cite{31}, which integrates their reward scores to estimate their contribution to answering the question, as shown in equation~\ref{ucb}. A greedy algorithm is employed, selecting the node with the highest score at each step. Specifically, the root node is initialized with the topic entity of the question, denoted as $Q_{original}$.
When multiple nodes produce equivalent UCB1 scores, a randomized selection protocol is activated to preserve exploration diversity. This selection process recursively explores the branching structure until it reaches a terminal leaf node, thereby determining the search trajectory.

\begin{equation}
UCB1_{i}=V_{i}+c\cdot\sqrt{\frac{2\cdot\ln N}{n_{i}}}\quad,
\label{ucb}
\end{equation}
where $V_{i}$ denotes the empirical average reward score of node $Leaf_{i}$, as generated by the LLM during the evaluation phase; $c$ is the exploration constant, set to 0.5 \cite{17,102}; $n_{i}$ represents the visit count of this node, and $N$ is the visit count of its parent node.

\subsection{Reflection and Evaluation Based on LLM}

\textbf{Reflection.} In this phase, a reflection mechanism \cite{32} is employed. The LLM first generates a brief reflection based on the current path $C^{path}_{i}$ to propose candidate next-hop nodes, denoted as $Leaf_{LLM1}$. Subsequently, $Leaf_{LLM1}$ is processed through equation \ref{coshopi} and \ref{leafhopsi} to determine the actual next-hop node $Leaf_{i+1}$. Finally, this reflection assesses whether appending $Leaf_{i+1}$ to $C^{path}_{i}$ is sufficient for answering the question. If so, the model outputs an \texttt{<end>} flag to terminate the path generation process. Otherwise, it continues by generating the next-hop node $Leaf_{LLM2}$. Through this iterative process, the LLM updates the reasoning path $C^{path}_{i+1}$, formed by concatenating $Leaf_{i+1}$ into $C^{path}_{i}$.
\begin{equation}
C^{path}_{i+1}=[Q_{original};Leaf_{0};Leaf_{1};...;Leaf_{i};Leaf_{i+1}].
\end{equation}

This evolving path $C^{path}_{i+1}$ serves as the context for subsequent LLM reasoning, guiding generation by reflecting prior steps and dynamically adjusting output diversity.

\textbf{Evaluation} The LLM assigns reward scores $V_{i+1}$ to newly generated nodes $Leaf_{i+1}$ based on three criteria: single-step correctness, semantic coherence, and node validity \cite{18}. Specifically, when erroneous steps emerge during intermediate stages, the corresponding node $Leaf_{i+1}$ receives a reduced reward score. This penalization mechanism discourages the selection of such branches, mitigating robustness degradation caused by mid-process errors. The scoring is formalized as follows:

\begin{equation}
V_{i+1}=\mathrm{LLM}(question:C^{path}_{i+1}, prompt),
\end{equation}
where $C^{path}_{i+1}$ denotes the accumulated path comprising all selected nodes up to the $i+1$-th iteration, and \textit{prompt} indicates the system prompt used to guide the LLM’s evaluation.

\subsection{Subgraph Expansion Based on Constraints}
During the previous phase, although the LLM generates new nodes $Leaf_{LLM1}$, directly using them as input often results in unstable performance. Therefore, node expansion is constrained by the following formulas.

\begin{equation}
CosHop_{i}=cos(Leaf_{LLM1}, LeafHop_{i}),
\label{coshopi}
\end{equation}

\begin{equation}
Leaf_{i+1} = \{\, v \in LeafHop_{i} \mid CosHop_{i}(v) > \textit{threshold} \,\},
\label{leafhopsi}
\end{equation}
where $LeafHop_{i}$ denotes the set of one-hop nodes of $Leaf_{i}$ in the KG; $CosHop_{i}$ represents the cosine similarity between each node in $LeafHop_{i}$ and node $Leaf_{LLM1}$. $Leaf_{i+1}$ refers to the subset of $LeafHop_{i}$ whose cosine similarity $CosHop_{i}$ exceeds the predefined threshold. The \textit{threshold} is set to 0.5.


\subsection{Reasoning Path Simulation}
Building upon the selection and expansion mechanisms, this stage uses a multi-criteria termination protocol to enhance exploration efficiency and maintain semantic integrity. This process maintains a path state tuple ($C^{path}_{i}$, $V_i$, $n_i$) and updates it during each simulation, where $V_i=\{V_0,...,V_i\}$ contains historical reward scores for the current path $C^{path}_{i}$, and $n_i$ records the visit count of the current node. Path simulation terminates when any of these conditions are met:
\begin{itemize}
\item \textit{Path Repetition}: The simulation terminates if the newly generated path is identical to the previous one.
\item \textit{Confidence Collapse}: The average of the evaluation scores falls below the threshold:
    \begin{equation}
    P_{end} = Cos(C^{path}_{i}, Q_{original}) < \gamma_{score}.
    \end{equation}
\item \textit{Structural Termination}: The expansion terminates when a leaf node is detected.
\item \textit{Solution Verification}: The LLM outputs an \texttt{<end>} flag after semantic validation.
\end{itemize}

The simulation integrates structured KG semantics through dual-channel processing: explicit path sequences $C^{path}_{i}$ guide the LLM’s reasoning, while implicit embedding similarity $CosHop_{i}$ ensures topological consistency. This method balances reasoning patterns with graph constraints, addressing both hallucination issues and computational efficiency requirements.

In the above process, we integrate the question and the existing path by serializing them. This serialization enables the reasoning path selected by MCTS to provide structured inference steps for the LLM. Meanwhile, it provides semantic support for subgraph extraction, ensuring that each inference step aligns with the actual relations in the KG.

\subsection{Backpropagation of Reward Score}

Backpropagation refines the selection phase by recursively propagating reward increments from leaf nodes to parent nodes through a bottom-up feedback mechanism. 
This upward propagation dynamically adjusts path weights in probabilistic models. The mathematical formulation is as follows: 

\begin{equation}
r_{N_{increment}} = \frac{\sum_{i=1}^{\text{branch}} Visits_{N_{children}} \cdot r_{N_{children}}}{\sum_{i=1}^{\text{branch}} Visits_{N_{children}}},
\end{equation}

\begin{equation}
Visits_{N_{children}} = Visits_{N_{children}} + 1,
\end{equation}

\begin{equation}
r_{parent} \leftarrow \alpha\cdot r_{parent}+(1-\alpha)\cdot r_{N_{children}},
\end{equation}
where $Visits_{N_{children}}$ denotes the number of visits to node $N_{children}$, which is used to balance the contribution of the frequently visited node to the increment. $r_{N_{increment}}$ denotes the increment in the parent node’s reward provided by the child node, and $\alpha$ is used to balance the original reward of the parent node and the increment from the child node. We set $\alpha$ to 0.7. These phases are repeated until the termination condition is met or the maximum tree depth is reached.

\section{Experiments}

In this phase, we conduct the experiments to address the following research questions: (1) Does our method improve the final QA performance? (2) Does our subgraph retrieval method generate smaller and higher-quality subgraphs? (3) Is each of the intermediate steps in the method effective?

\subsection{Experimental Settings}
\subsubsection{Datasets}

We evaluate our method on two KBQA datasets based on Freebase and one dataset based on the WikiMovies domain. A summary of dataset statistics is provided in Table~\ref{DATASET_INFO1}.
Freebase is a large open source KB composed of structured data, comprising approximately 68 million entities and 1 billion relations.
WebQuestionsSP (WebQSP) \cite{35}, derived from the WebQuestions dataset \cite{36}, augments natural language questions with SPARQL annotations and evaluates answers by executing the corresponding queries on Freebase. This enables a more rigorous assessment of whether a question can be answered. It contains 4,737 answerable natural language questions and splits the train/dev/test sets as 2,848/250/1,639.
Complex WebQuestions 1.1 (CWQ) \cite{37}, based on WebQSP, introduces questions with complex structures such as conjunctions, superlatives, comparatives, and compositions. The dataset consists of 34,689 cases, including 27,639 training instances, 3,531 test instances, and 3,519 validation cases. 
MetaQA \cite{96} is a large-scale multi-hop QA dataset constructed from a movie-domain knowledge graph. It includes over 400k questions across 1-hop, 2-hop, and 3-hop settings. Across all three subsets, MetaQA presents a high proportion of complex multi-hop queries, making it a suitable testbed for assessing model robustness beyond Freebase.

\begin{table}[]
    \centering
    \caption{Statistics of the datasets used in this paper.}
    \begin{tabular}{@{}ccccc@{}}
    \toprule
    \textbf{Dataset} & \textbf{KG} & \textbf{\#Train} & \textbf{\#Validation} & \textbf{\#Test} \\\midrule
    WebQSP          &Freebase & 2,848            & 250                   & 1,639   \\
    CWQ             &Freebase & 27,639           & 3,519                 & 3,531   \\
    MetaQA 1-hop & WikiMovies & 96,106 & 9,992 & 9,947 \\
    MetaQA 2-hop & WikiMovies & 118,980 & 14,872 & 14,872 \\
    MetaQA 3-hop & WikiMovies & 114,196 & 14,274 & 14,274 \\ \bottomrule
    \end{tabular}
    \label{DATASET_INFO1} 
\end{table}

\subsubsection{Evaluation Metrics}

In QA tasks, our methodology employs two complementary metrics to assess model performance. Since a question may have multiple correct answers, we use Hits@k to evaluate retrieval performance. Hits@k measures the percentage of questions for which at least one correct answer appears in the top-k retrieved answers. This serves as an upper bound indicator of the model’s ability to retrieve relevant answers within limited search conditions. To complement this retrieval-focused evaluation, the F1-score is used to capture the trade-off between precision and recall in answer selection.

The evaluation framework also incorporates a cost-effectiveness analysis \cite{39}, assessing the trade-off between generation quality and computational cost. Generation quality is measured using task-specific metrics, such as Hits@1, which evaluates the model's precision in ranking the most relevant answer at the top. Computational cost is measured by the overall runtime of our method. Cost-effectiveness is then quantified as the ratio of Hits@1 to overall runtime, enabling direct comparison of how different models balance predictive accuracy against computational demands.

\subsubsection{Implementation Details}
For our method, the reasoning depth is set to 3 for both the WebQSP and MetaQA datasets, and to 4 for the CWQ dataset. The number of MCTS simulations is set to 5. For baseline methods, we employ DeepSeek-V3 as the LLM backbone due to its strong QA performance and relatively low computational cost \cite{105,106}. In addition, all experiments are conducted on 2 NVIDIA RTX 4090 24GB GPUs.

\subsubsection{Baseline Models}

We evaluate our model against representative baselines from three categories. For IR-based models, we consider Key-Value Memory Network (KV-MemNN) \cite{40}, EmbedKGQA \cite{41}, GRAFT-Net \cite{24}, NSM \cite{6}, PullNet \cite{5}, NSM\textsubscript{+p} \cite{6}, and SR \cite{44}. For SP-based models, we include SPARQA \cite{45}, TextRay \cite{61}, MULTIQUE \cite{46}, SRGCN \cite{47}, and Relational Chain-based Embedded KGQA (Rce-KGQA) \cite{48}. For LLM-based methods, we compare with ToG \cite{50}, GoG \cite{49}, StructGPT \cite{53}, ChatGPT \cite{90}, Flan-T5-XL \cite{54}, FlexKBQA \cite{83}, KBQA-o1 \cite{86}, and MCTS-KBQA \cite{88}.
 
\subsection{Overall QA Evaluation}

We compare the performance of M-ER against previous state-of-the-art methods. Table~\ref{Overall_table} reports the results on the WebQSP and CWQ datasets, while Table~\ref{Overall_table1} reports the performance on the MetaQA dataset.

\begin{table}[] 
    \centering
    \caption{Overall QA evaluation of WebQSP and CWQ datasets. Hits@1 (\%) and F1-score are reported. $*$ refers to the reproduced result from our work. \textbf{Bold} numbers denote the best performance, and \underline{underlined} numbers denote the second-best performance.}
    \scalebox{1.0}{
    \begin{tabular}{ccccc}
    \hline
    \multirow{2}{*}{Models}     & \multicolumn{2}{c}{WebQSP} & \multicolumn{2}{c}{CWQ}  \\ \cline{2-5}
                               & Hits@1         & F1-score  & Hits@1   & F1-score       \\\hline
    \multicolumn{5}{c}{SP-based   models}                            \\\hline
    SPARQA\cite{45}        & -            & -           & 31.57       & -     \\
    TextRay\cite{61}      & 72.16           & 60.3        & 40.83       & 33.87   \\
    MULTIQUE\cite{46}        & 69.84         & 61.16        & 37.61       & 31.24   \\
    SRGCN\cite{47}             & -            & 56.6        & -          & -    \\
    Rce-KGQA\cite{48}            & 70.4         & -           & 48.3          & -       \\\hline
    \multicolumn{5}{c}{IR-based   models}                              \\\hline
    KV-Mem\cite{40}        & 46.6         & 34.5        & 18.4       & 15.7   \\
    EmbedKGQA\cite{41}       & 66.6         & -  & 32.0       & -   \\
    GRAFT-Net\cite{24}        & 68.7         & 62.3        & 36.8       & 32.7   \\
    NSM\textsuperscript{*}\cite{6}  & 67.1  & 63.4    & 47.2       & 41.9   \\
    PullNet\cite{5}          & 68.1         & -    & 45.9       &-  \\
    NSM\textsubscript{+p}\cite{6}     & 71.5         & 64.2        & 47.9       & 42.4  \\
    SR\textsuperscript{*}\cite{44} & 71.8  & 65.4     & 52.8    & 45.8  \\ \hline
    \multicolumn{5}{c}{LLM-based}                       \\\hline
    GoG\textsuperscript{*}\cite{49} & 76.2   & 66.5   & 54.3       & 50.1     \\
    ToG\textsuperscript{*}\cite{50}     & \underline{77.2}    & \underline{70.9}    & 49.8  & 46.5     \\
    StructGPT\textsuperscript{*}\cite{53}     & 73.3   & 69.5    & -    & - \\
    ChatGPT\cite{90}    & 66.8   & -    & 39.9    & - \\
    Flan-T5-XL\cite{54} & 31.0   & -   & 14.7 & - \\
    FlexKBQA\cite{83}    & - & 60.6 & - & - \\ 
    KBQA-o1\cite{86}     & - & 67.0 & - & - \\ 
    MCTS-KBQA\textsuperscript{*}\cite{88} & 74.6 & 65.4 & \underline{68.3} & \underline{64.2} \\ \hline
    \textbf{Our method}  & \textbf{78.88}  & \textbf{72.17} & \textbf{73.46} & \textbf{65.28}   \\\hline
    \end{tabular}
    }
    \label{Overall_table} 
\end{table}

\begin{table}[t] 
    \centering
    \caption{Comparison of Hits@1 (\%) across state-of-the-art methods on the MetaQA dataset. $*$ indicates that the results of the corresponding method are reproduced by our work under the same experimental settings. The best performance in each column is indicated in \textbf{bold}, while the second-best performance is \underline{underlined}.}
    \scalebox{0.8}{
    \begin{tabular}{ccccc}
    \hline
    \multirow{1}{*}{Models} & MetaQA 1-hop & MetaQA 2-hop & MetaQA 3-hop & Average \\ \hline
    Rce-KGQA\cite{48}            & \underline{98.3} & \underline{99.7} & 97.9 & 98.6 \\
    KV-Mem\cite{40}              & 96.2 & 82.7 & 48.9 & 75.9 \\
    EmbedKGQA\cite{41}          & 97.5 & 98.8 & 94.8 & 97.0 \\
    GRAFT-Net\cite{24}           & 97.0 & 94.8 & 77.7 & 89.8 \\
    NSM\textsuperscript{*}\cite{6}       & 97.3 & 99.6 & 98.2 & 98.4 \\
    PullNet\cite{5}             & 97.0 & \textbf{99.9} & 91.4 & 96.1 \\
    NSM\textsubscript{+p}\cite{6}  & 97.3 & \textbf{99.9} & \underline{98.9} & \underline{98.7} \\
    StructGPT\textsuperscript{*}\cite{53} & 94.7 & 94.3 & 79.6 & 89.5 \\
    \textbf{Our method} & \textbf{98.6} & \underline{99.7} & \textbf{99.3} & \textbf{99.2} \\ \hline
    \end{tabular}
    }
    \label{Overall_table1} 
\end{table}

For the WebQSP and CWQ datasets, the results in Table \ref{Overall_table} demonstrate that our method achieves strong performance across both benchmarks. On the WebQSP dataset, our framework obtains 78.88\% Hits@1 and 72.17 F1-score, significantly outperforming all baseline models. Compared to the SP-based method, Rce-KGQA, our approach improves Hits@1 by 8.48\%. 
On the CWQ dataset, M-ER achieves 73.46\% Hits@1 and 65.28 F1-score, surpassing IR-based and SP-based methods. For example, compared to the best IR-based model, SR, our method yields gains of 20.66\% in Hits@1 and 19.48 in F1-score on the CWQ dataset. 
While LLM-based approaches, such as GoG, ToG, StructGPT, and MCTS-KBQA, show competitive performance, our method still achieves the best results. Notably, it surpasses GoG by 2.68\% in Hits@1 on the WebQSP dataset, and by 19.16\% in Hits@1 on the CWQ dataset. Similarly, compared with MCTS-KBQA, the Hits@1 metric increases by 4.28\% and 5.16\% on the two datasets, respectively. 
These consistent improvements highlight the robustness and generalization of our framework across different datasets.

For the MetaQA dataset, M-ER achieves the highest average performance, with an average Hits@1 score of 99.2\%, surpassing all other methods. Specifically, M-ER obtains Hits@1 scores of 98.6\%, 99.7\%, and 99.3\% for 1-hop, 2-hop, and 3-hop datasets, respectively. It outperforms PullNet on the 1-hop (97.0\%) and 3-hop (91.4\%) datasets, while slightly underperforming compared to PullNet and NSM\textsubscript{+p} on the 2-hop dataset (99.9\%). These results highlight M-ER's ability to generalize across knowledge graphs from different domains, suggesting that the observed performance gains are not limited to Freebase.

\textit{Summary} Our M-ER achieves superior performance on the WebQSP, CWQ, and MetaQA datasets, with significant advantages in multi-hop KBQA.  These improvements stem from the integration of a key node selection mechanism that focuses on answer relevance.

\begin{figure}
    \centering
    \includegraphics[width=1\linewidth]{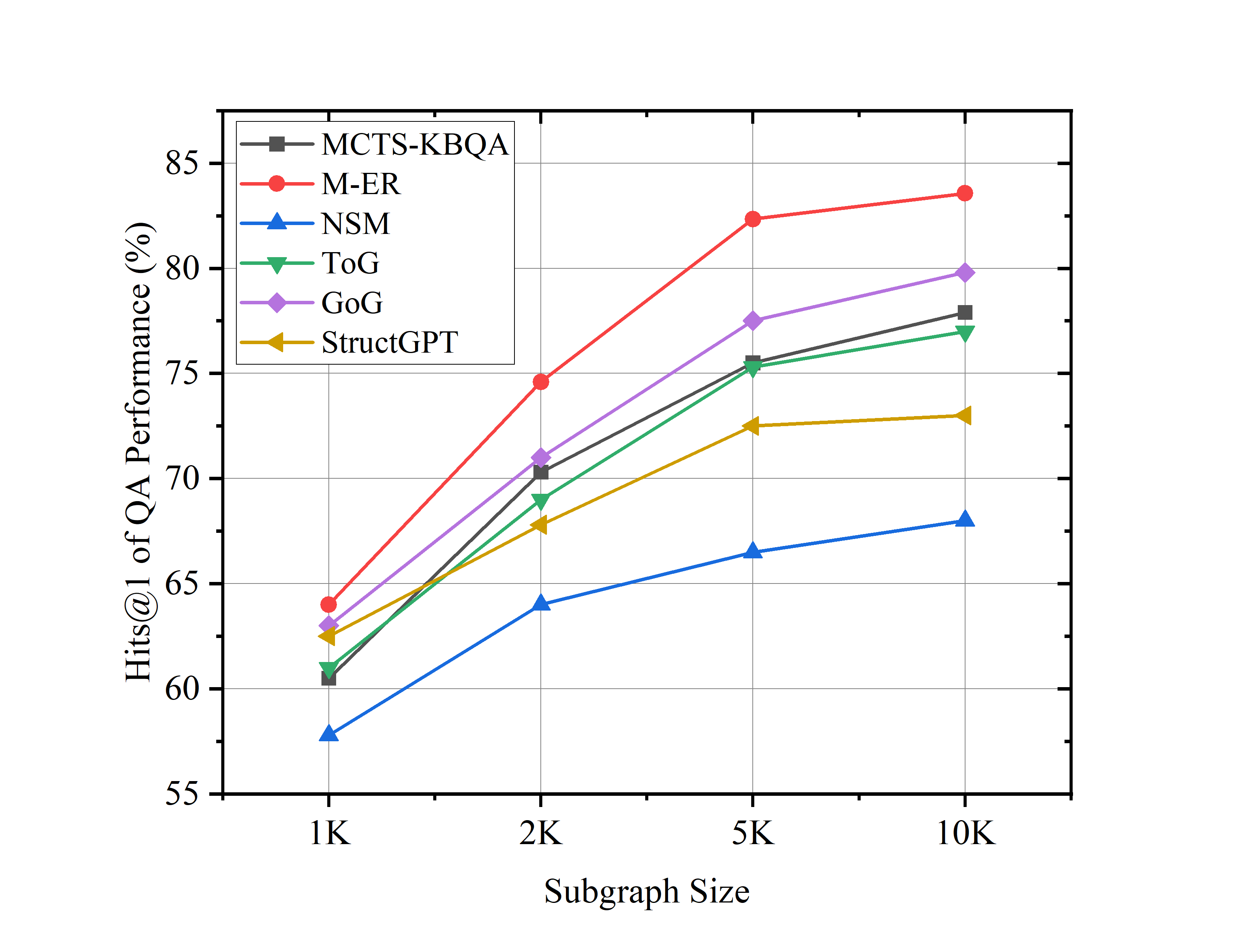}
    \caption{The impact of different subgraph sizes on the WebQSP dataset for the Hits@1 metric. The baselines for comparison are MCTS-KBQA, NSM, ToG, GoG, StructGPT, and M-ER.}
    \label{fig:Hits@1}
\end{figure}

\begin{figure}
    \centering
    \includegraphics[width=1\linewidth]{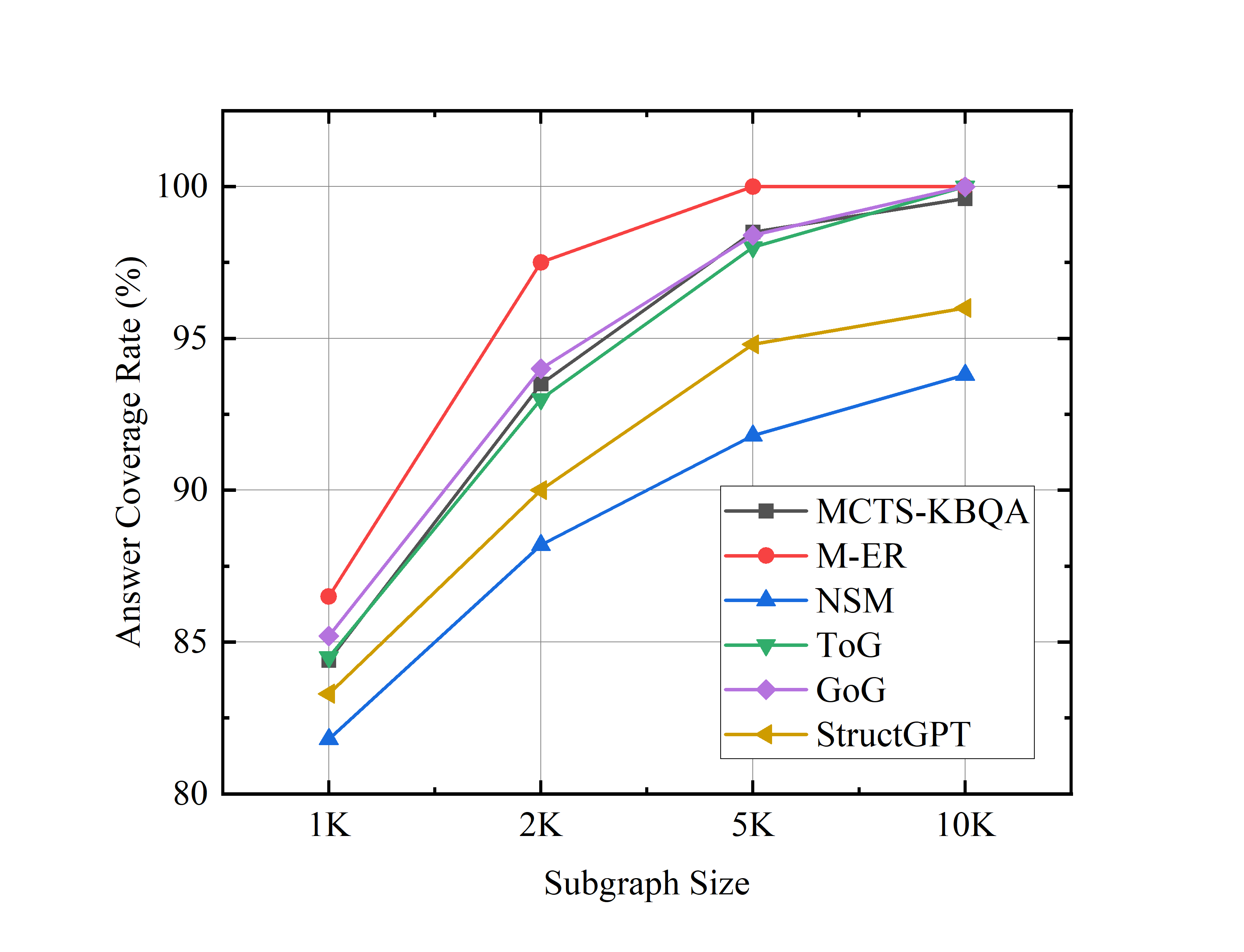}
    \caption{The impact of different subgraph sizes on the WebQSP dataset for the answer coverage rate. The baselines for comparison are MCTS-KBQA, NSM, ToG, GoG, StructGPT, and M-ER.}
    \label{fig:answeragerage}
\end{figure}

\subsection{Evaluation of Retrieved Subgraph}

As shown in Figure \ref{fig:Hits@1}, the experimental results clearly demonstrate the effectiveness of M-ER in terms of the Hits@1 metric across different subgraph sizes on the WebQSP dataset. M-ER demonstrates a more pronounced performance improvement as the subgraph size increases, rapidly surpassing all baseline methods, including MCTS-KBQA \cite{88}, NSM \cite{6}, ToG \cite{50}, GoG \cite{49}, and StructGPT \cite{53}. Notably, M-ER achieves over 80\% Hits@1 even with a subgraph size of only 5K, showcasing its capability to identify high-quality answers within smaller search spaces.

Figure \ref{fig:answeragerage}, which presents the answer coverage rate across different subgraph sizes, further validates the effectiveness of M-ER. It consistently outperforms all baselines, reaching near-perfect coverage at 5K and maintaining 100\% coverage at 10K, while other models require larger subgraphs to achieve comparable performance. These results underscore the robustness of M-ER in identifying correct reasoning paths while preserving essential semantic and structural information during subgraph pruning.

These empirical trends collectively reflect the superior scalability and adaptability of the M-ER framework. Compared with conventional models, M-ER benefits from its hybrid architecture that integrates MCTS with a node scoring mechanism. This synergy allows the model to effectively suppress irrelevant branches while favoring nodes that are semantically rich and relevant to the question. As a result, M-ER scales to larger graphs and more complex queries, making it particularly well-suited for real-world QA over large-scale knowledge graphs.

\textit{Summary} Our method significantly outperforms MCTS-KBQA, NSM, ToG, and other baselines in both Hits@1 and answer coverage rate, achieving superior performance with smaller subgraph sizes. These results demonstrate the effectiveness and efficiency of M-ER in enhancing QA performance while reducing computational overhead.

\subsection{Cost-Effectiveness Analysis}

The resource consumption of our method is primarily influenced by two factors: the use of LLMs and the design of the MCTS strategy. To evaluate the impact of these two components on computational resources, we design three experiments on the WebQSP dataset and report the wall-clock GPU hours, overall runtime, input/output token counts, and cost-effectiveness. The three experimental settings are summarized as follows.
\begin{itemize}
    \item \textbf{Non-LLM Setting.} 
    When LLMs are not involved, this experiment evaluates the wall-clock GPU hours, overall runtime, cost-effectiveness, and Hits@1 of M-ER (w/o LLM) by comparing it with classic IR-based KBQA methods, including SR~\cite{44}, NSM~\cite{6}, and DPR~\cite{98} + Reranker~\cite{99}, as shown in Table \ref{first}.
    \item \textbf{LLM-based Setting.} 
    When LLMs are involved, this experiment evaluates overall runtime, input/output token counts, cost-effectiveness, and Hits@1 of M-ER (w/ LLM) by comparing it with SOTA LLM-driven KBQA methods, such as ToG~\cite{50}, GoG~\cite{49}, and MCTS-KBQA~\cite{88}, as shown in Table \ref{second}. 
    \item \textbf{Search Strategy Comparison Setting.} 
    This experiment evaluates overall runtime, input/output token counts, cost-effectiveness, and Hits@1 by analyzing the impact of different search algorithms. Specifically, M-ER replaces MCTS with alternative search strategies, including ReAct~\cite{89}, depth-first search (DFS)~\cite{97}, and breadth-first search (BFS)~\cite{27}, as shown in Table \ref{third}. 
\end{itemize}

In the first set of experiments (Non-LLM Setting), we evaluate the wall-clock GPU hours, overall runtime, cost-effectiveness, and Hits@1 of different methods, as shown in Table \ref{first}. The results indicate that M-ER (w/o LLM) achieves the best performance across all metrics compared to the baselines. Specifically, M-ER attains the highest Hits@1 (72.37\%) and cost-effectiveness (44.13), while also exhibiting the lowest wall-clock GPU time (0.74s) and overall runtime (1.64s). In comparison, SR \cite{44} requires 1.42s GPU time and 2.05s overall runtime, NSM~\cite{6} takes 1.68s / 2.57s, and DPR~\cite{98} + Reranker~\cite{99} requires 0.98s / 1.94s. The superior performance of M-ER can be attributed to its search strategy. SR~\cite{44} selects subgraphs based on nodes with the highest similarity scores, which often leads to redundant expansions on large-scale knowledge graphs. In contrast, M-ER avoids unnecessary exploration while retaining question-relevant nodes. This strategy effectively reduces computational overhead and improves answer accuracy.

In the second set of experiments (LLM-based Setting), we evaluate the overall runtime, input/output token counts, cost-effectiveness, and Hits@1 by comparing our method with other approaches, as shown in Table~\ref{second}. The results indicate that M-ER (w/ LLM) achieves the best overall performance across most metrics. Specifically, M-ER attains the highest Hits@1 (78.88\%) while exhibiting the lowest overall runtime (6.08s). In terms of token usage, M-ER consumes 706 input tokens and 320 output tokens per question, which is comparable to ToG~\cite{50} (679/326 tokens) and GoG~\cite{49} (548/406 tokens). Despite a comparable token budget, M-ER achieves superior performance, as reflected by its higher Hits@1 and cost-effectiveness than other baselines. In contrast, MCTS-KBQA~\cite{88} shows lower accuracy (74.68\%) and cost-effectiveness (7.49) while consuming more tokens, indicating inefficient utilization of LLM calls. The superior performance of M-ER can be attributed to its search strategy and guidance mechanism. Although MCTS-KBQA~\cite{88} also uses MCTS, it lacks explicit guidance for selecting subsequent search directions and does not fully exploit the structural information within subgraphs. As a result, it often terminates at intermediate states as final answers, leading to lower accuracy and unstable performance.

In the third set of experiments (Search Strategy Comparison Setting), we evaluate overall runtime, input/output token counts, cost-effectiveness, and Hits@1 by comparing M-ER with different search strategies, as shown in Table \ref{third}. The results indicate that M-ER achieves the best overall performance across all metrics compared with other baselines. Specifically, M-ER attains the highest Hits@1 while also exhibiting the lowest overall runtime, resulting in the highest cost-effectiveness score.  In comparison, ReAct \cite{89} achieves a Hits@1 score of 73.23\% and 10.38s overall runtime, while BFS \cite{27} and DFS \cite{97} show slightly higher overall runtime and lower accuracy than our method. Regarding LLM token consumption, M-ER requires fewer tokens than ReAct \cite{89} (982/735), BFS \cite{27} (904/382), and DFS \cite{97} (857/356).  Overall, these results demonstrate that integrating MCTS with LLMs in M-ER not only improves answer accuracy but also ensures more efficient utilization of computational resources.

\textit{Summary} The cost-effectiveness analysis demonstrates that M-ER consistently outperforms other methods across all three experimental setups, achieving a lower overall runtime. These results further confirm the efficiency of our method.

\begin{table}[h]
    \centering
    \caption{The wall-clock GPU hours, overall runtime, cost-effectiveness, and Hits@1 of M-ER under the non-LLM setting on the WebQSP dataset. \textbf{Bold} numbers denote the best performance, and \underline{underlined} numbers denote the second-best performance.}
    \scalebox{0.7}{
    \begin{tabular}{ccccc}
    \hline
    Method & Hits@1 & Wall-clock GPU Time & Overall Runtime  & Cost-Effectiveness \\ \hline
    SR\cite{44}       & \underline{71.83} & 1.42 & 2.05   & \underline{35.04}  \\
    NSM\cite{6}     & 67.18   & 1.68 & 2.57   & 26.14 \\ 
    DPR\cite{98} + Reranker\cite{99} & 65.76 & \underline{0.98} & \underline{1.94} & 33.90 \\
    M-ER (w/o LLM)  &\textbf{72.37} &\textbf{0.74} &\textbf{1.64}   & \textbf{44.13}  \\\hline
    \end{tabular}
    }
    \label{first}
\end{table}

\begin{table}[]
    \centering
    \caption{The overall runtime, input/output token counts, cost-effectiveness, and Hits@1 of M-ER under the LLM-based setting on the WebQSP dataset. \textbf{Bold} numbers denote the best performance, and \underline{underlined} numbers denote the second-best performance.}
    \scalebox{0.7}{
    \begin{tabular}{ccccc}
    \hline
    Method & Hits@1 &  Overall Runtime  & Input Tokens / Output Tokens & Cost-Effectiveness \\ \hline
    ToG\cite{50}  & \underline{77.26}  & \underline{9.33} & \underline{679}/\underline{326}  & \underline{8.28} \\
    GoG\cite{49}  & 76.24  & 9.79 & \textbf{548}/406  & 7.79 \\
    MCTS-KBQA\cite{88} & 74.68 & 9.97 & 892/629  & 7.49 \\ 
    M-ER (w/ LLM) &\textbf{78.88}  &\textbf{6.08} & 706/\textbf{320}  & \textbf{12.97}\\\hline
    \end{tabular}
    }
    \label{second}
\end{table}
 
\begin{table}[H]
    \centering
    \caption{The overall runtime, input/output token counts, cost-effectiveness, and Hits@1 of M-ER under the search strategy comparison setting on the WebQSP dataset. \textbf{Bold} numbers denote the best performance, and \underline{underlined} numbers denote the second-best performance.}
    \scalebox{0.74}{
    \begin{tabular}{ccccc}
    \hline
    Method & Hits@1  & Overall Runtime  & Input Tokens / Output Tokens & Cost-Effectiveness \\ \hline
    ReAct\cite{89} & 73.23 & \underline{10.38} & 982/735  & \underline{7.05} \\
    BFS\cite{27} &74.17  &13.15   & 904/382  &5.64 \\
    DFS\cite{97} &\underline{75.24} &13.06   & \underline{857}/\underline{356}  &5.76 \\
    M-ER &\textbf{78.88} &\textbf{6.08} & \textbf{706/320} &\textbf{12.97} \\ \hline
    \end{tabular}
    }
    \label{third}
\end{table}

\subsection{Ablation Study}

To evaluate the impact of individual components in M-ER, we conduct ablation studies under the following settings:
\begin{itemize}
\item W/o Reflection: The reflection of the LLM is removed, and the subgraph is directly extended based on the similarity between the question and nodes.
\item W/o Evaluation: The evaluation step is omitted, resulting in random node selection during each expansion step.
\item W/o MCTS: Without MCTS, node search is based on similarity, consistent with SR~\cite{44}.
\item W/o LLM: This setting removes both the reflection and evaluation mechanisms of the LLM.
\end{itemize}

As can be seen from Table~\ref{Ablation study}, removing the LLM module has the greatest impact on model performance, with a 5.37\% decrease in Hits@1. This indicates that the reflection and evaluation mechanisms of the LLM play a crucial role in improving overall performance. Specifically, the reflection mechanism directs the next search nodes based on the current subgraph, while the evaluation mechanism scores nodes in the current subgraph and feeds them back to the node selection stage.  
The removal of the reflection mechanism reduces Hits@1 by 1.86\%, indicating that the mechanism also plays a significant role in guiding the subsequent search process.
The removal of the MCTS module leads to a 4.59\% performance degradation, indicating that MCTS effectively reduces redundant exploration by constraining the topological paths between topic entities and answer entities in the knowledge base. 
Finally, removing the evaluation module reduces Hits@1 by 1.2\%. By scoring the generated nodes at each step, the evaluation mechanism more accurately captures semantic dependencies relevant to the question.

\textit{Summary} The ablation study demonstrates that integrating these modules reduces unnecessary reasoning steps, resulting in better overall performance compared to the ablated configurations.

\begin{table}[t]
    \centering
    \caption{Ablation study on the WebQSP dataset. The best performance in each column is indicated in \textbf{bold}, while the second-best performance is \underline{underlined}.}
    \begin{tabular}{cccc}
    \hline
    Model            & Hits@1 & Hits@10 & F1-score      \\\hline
    W/o Reflection & 77.02 & 81.64 & 67.24  \\
    W/o Evaluation & \underline{77.68} & \underline{83.85} & \underline{69.96}    \\
    W/o MCTS       & 74.29 & 79.03 & 66.88   \\
    W/o LLM       & 73.51 & 77.21 & 64.97   \\
    Our method       & \textbf{78.88} & \textbf{85.21} & \textbf{72.17}  \\\hline
    \end{tabular}
    \label{Ablation study}
\end{table}

\subsection{Performance on Different LLMs}

To investigate the impact of different LLMs, we compare the performance of six models on the WebQSP dataset, including llama3.1-8B, Qwen2-7B, DeepSeek-R1-Distill-Qwen-32B, GPT-3.5, GPT-4o, and DeepSeek-V3, as shown in Table~\ref{Performance on different LLM}. 
GPT-4o achieves the best Hits@1 score of 80.96\% and ranks second on Hits@10 (84.49\%) and F1-score (71.32). DeepSeek-V3 attains the highest Hits@10 (85.21\%) and F1-score (72.17), while achieving a competitive Hits@1 of 78.88\%. GPT-3.5 ranks second in Hits@1 (79.52\%).
DeepSeek-R1-Distill-Qwen-32B achieves 72.95\% Hits@1, 74.08\% Hits@10, and 67.69 F1-score, outperforming the smaller open-source models but remaining below GPT-3.5, GPT-4o, and DeepSeek-V3. In contrast, Qwen2-7B and llama3.1-8B exhibit lower overall performance. Specifically, Qwen2-7B achieves 70.68\% Hits@1, 71.98\% Hits@10, and 65.02 F1-score, while llama3.1-8B obtains 68.23\% Hits@1, 69.21\% Hits@10, and 64.03 F1-score.
Overall, larger and more advanced models (e.g., GPT-4o and DeepSeek-V3) demonstrate clear advantages across multiple metrics, indicating stronger multi-hop reasoning capabilities. In contrast, smaller models show consistent performance degradation across all metrics, suggesting limitations in complex reasoning and semantic consistency.

\textit{Summary} The comparative analysis highlights performance variations across different models. While smaller models, such as llama3.1-8B, achieve reasonable performance, DeepSeek-V3 demonstrates an excellent ability to align semantic dependencies. 

\begin{table}[H]
    \centering
    \caption{Performance on different LLMs on the WebQSP dataset. The best performance in each column is indicated in \textbf{bold}, while the second-best performance is \underline{underlined}.}
    \begin{tabular}{cccc}
    \hline
    \multicolumn{1}{c}{Model}                        & \multicolumn{1}{c}{Hits@1} & \multicolumn{1}{c}{Hits@10} & \multicolumn{1}{c}{F1-score} \\\hline
    \multicolumn{1}{c}{llama3.1-8B}                  & \multicolumn{1}{c}{68.23} & \multicolumn{1}{c}{69.21}  & \multicolumn{1}{c}{64.03} \\
    \multicolumn{1}{c}{Qwen2-7B}                     & \multicolumn{1}{c}{70.68} & \multicolumn{1}{c}{71.98}  & \multicolumn{1}{c}{65.02} \\
    \multicolumn{1}{c}{DeepSeek-R1-Distill-Qwen-32B} & 72.95 & 74.08  & 67.69  \\   
    \multicolumn{1}{c}{GPT-3.5}                       & \underline{79.52} & 81.48  & 68.65  \\ 
    \multicolumn{1}{c}{GPT-4o}                       & \textbf{80.96} & \underline{84.49}  & \underline{71.32}  \\   
    \multicolumn{1}{c}{DeepSeek-V3}                       & 78.88 & \textbf{85.21}  & \textbf{72.17}  \\ \hline
    \end{tabular}
    \label{Performance on different LLM}
\end{table}

\subsection{Case Study}
\begin{figure*}[t] 
\centering
\includegraphics[scale=0.9]{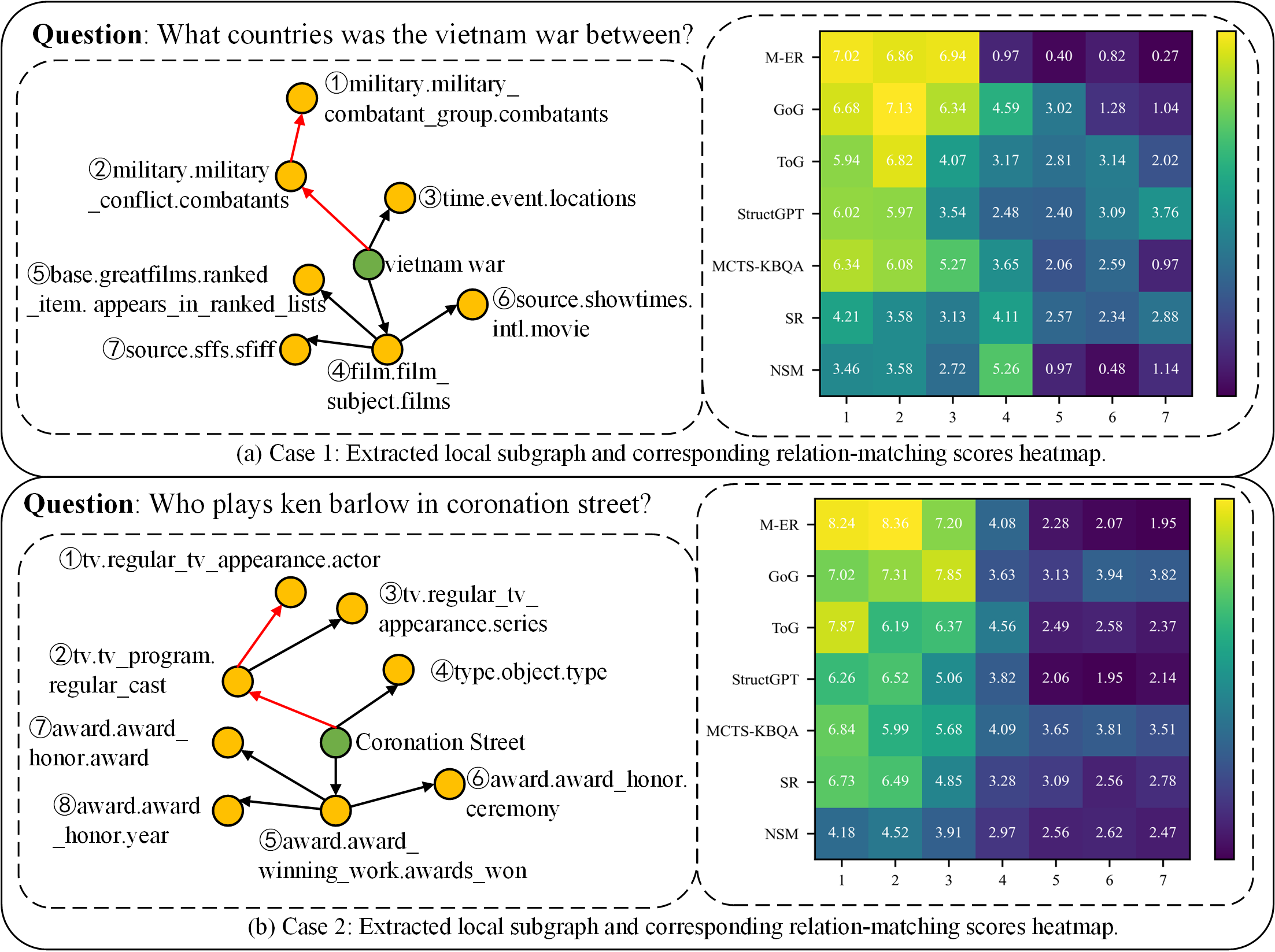}
\caption{Illustration of the extracted local subgraph and the corresponding relation-matching score heatmap, where the numbers denote relation indices. The red arrows indicate the correct reasoning path.}
\label{case_all}
\end{figure*}

Subgraphs, as effective path combinations between entities, provide more focused insight than a complete KB by revealing the reasoning behind each target link. To further illustrate the advantages of our method, we conduct two case studies on the WebQSP dataset to evaluate whether it assigns higher relation–question similarity scores and better captures relevant contextual information, as shown in Figure~\ref{case_all}. For each case, we present the extracted local subgraph and report the relation-matching scores—computed by our method and several representative baselines—between the question and each candidate relation.

Across both cases in Figure~\ref{case_all}, M-ER exhibits a clear separation between contextually relevant and irrelevant relations in its similarity scores. This pattern suggests that M-ER captures the semantic dependencies between the question and candidate relations more accurately than existing baseline methods. Case 2 illustrates this contrast most clearly. M-ER produces sharp peaks for the correct relations and yields the largest gap between high and low similarity scores. In particular, it assigns high scores to the role-related relations (relations 1-3), with relation 1 reaching a score of 8.24, while assigning low scores to the award-related relations (relations 5–8), with relation 8 decreasing to 1.95. In comparison, StructGPT~\cite{53} and MCTS-KBQA~\cite{88} exhibit much flatter score distributions, which makes it difficult to distinguish relevant relations from irrelevant ones. NSM~\cite{6} shows an even smaller margin, with only a difference of 2.05 between its highest and lowest scores. Case 1 demonstrates a similar overall trend. Overall, these two cases demonstrate that M-ER yields more discriminative similarity patterns and clearer semantic separation between relevant and irrelevant relations. This enhanced discrimination facilitates more accurate identification of relevant relations, resulting in higher-quality subgraphs and better question-answering performance.

\textit{Summary} These case studies demonstrate that our method improves semantic discrimination in subgraph extraction. Compared to SR and other baselines, such as GoG, ToG, StructGPT, and MCTS-KBQA, our approach more effectively identifies contextually relevant nodes while suppressing irrelevant ones, thus enhancing the quality of the subgraph.

\begin{figure*}[]
    \centering
    \includegraphics[width=1\linewidth]{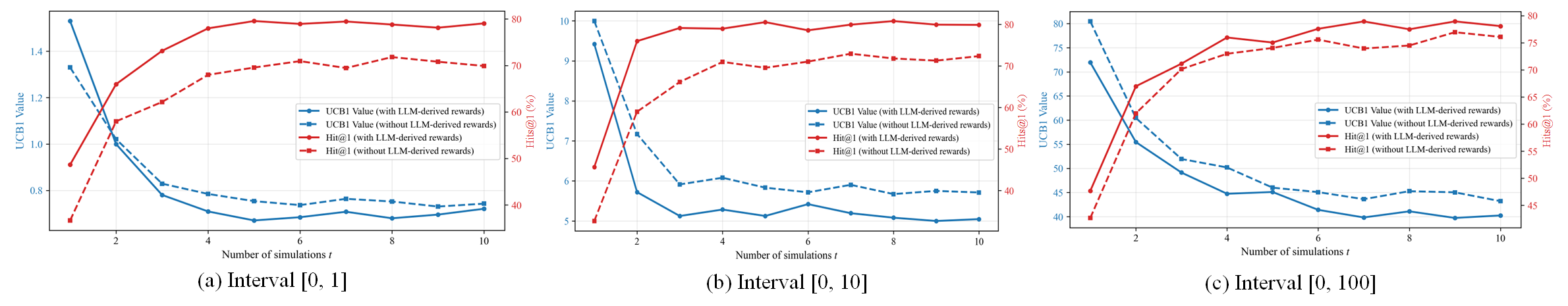}
    \caption{Empirical convergence of UCB1 and Hits@1 over 10 MCTS iterations under different reward ranges on the WebQSP dataset for two methods (with LLM-derived rewards and without LLM-derived rewards). The (a)–(c) correspond to the results for reward ranges of [0, 1], [0, 10], and [0, 100], respectively. The UCB1 curve (blue) starts from its upper bound and decreases toward the stable value. Simultaneously, Hits@1 (red) curve plateaus.}
    \label{responseucb-webqsp}
\end{figure*}
\vspace{2pt}
\begin{figure*}[]
    \centering
    \includegraphics[width=1\linewidth]{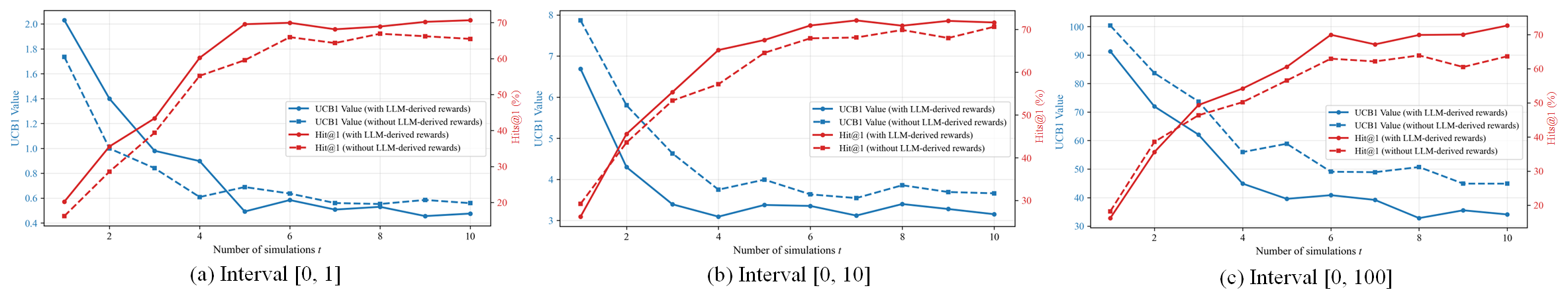}
    \caption{Empirical convergence of UCB1 and Hits@1 over 10 MCTS iterations under different reward intervals on the CWQ dataset for two methods (with LLM-derived rewards and without LLM-derived rewards). The (a)–(c) correspond to the results for reward ranges of [0, 1], [0, 10], and [0, 100], respectively. The UCB1 curve (blue) starts from its upper bound and decreases toward the stable value. Simultaneously, Hits@1 (red) curve plateaus.}
    \label{responseucb-cwq}
\end{figure*}
\vspace{2pt}
\begin{figure*}[]
    \centering
    \includegraphics[width=1\linewidth]{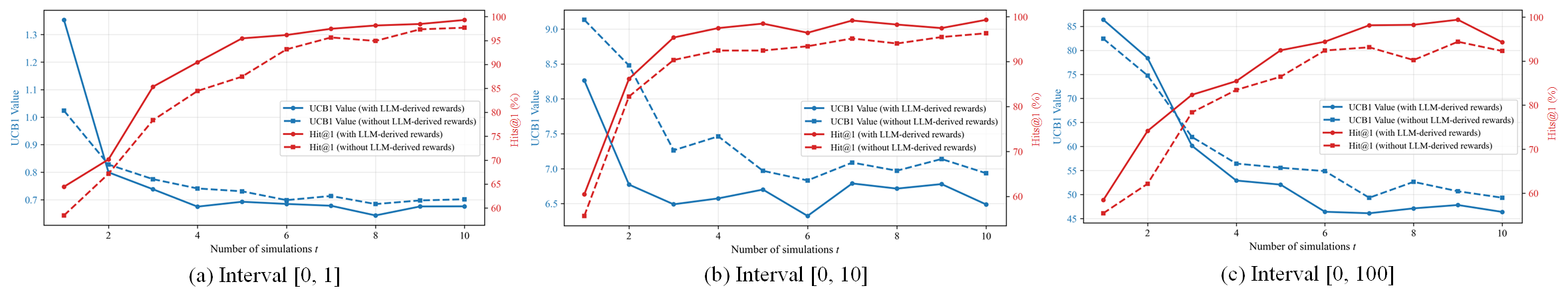}
    \caption{Empirical convergence of UCB1 and Hits@1 over 10 MCTS iterations under different reward intervals on the MetaQA-3hop dataset for two methods (with LLM-derived rewards and without LLM-derived rewards). The (a)–(c) correspond to the results for reward ranges of [0, 1], [0, 10], and [0, 100], respectively. The UCB1 curve (blue) starts from its upper bound and decreases toward the stable value. Simultaneously, Hits@1 (red) curve plateaus.}
    \label{responseucb-meta}
\end{figure*}

\subsection{Discussion}
\subsubsection{Empirical Analysis of UCB1 Convergence}
To analyze the empirical convergence of UCB1 \cite{31} and the iteration count required for convergence, we evaluate both the convergence curves of UCB1 and Hits@1 on three benchmark datasets (WebQSP, CWQ, and MetaQA-3hop datasets). The experiments are conducted under two conditions, with LLM-derived rewards and without LLM-derived rewards, to analyze the impact of LLM-derived rewards on convergence. Specifically, the range of LLM-derived rewards is treated as a hyperparameter, since its scale directly affects the exploration–exploitation balance in UCB1, thereby influencing convergence speed and overall performance. Following the configurations adopted in prior work, we evaluate three representative reward ranges, including [0, 1] \cite{17}, [0, 10] \cite{18,103}, and [0, 100] \cite{2,104}. To ensure a fair comparison across different methods, all other parameters are kept unchanged \cite{17,102}. For each reward range ([0,1], [0,10], [0,100]), we use the same set of \(5\) randomly sampled points. In each iteration, the UCB1 values of these points are averaged to produce the UCB1 value for that iteration. This process is repeated for 10 iterations to generate the UCB1 convergence curve for each reward range. The corresponding results are reported in Figures~\ref{responseucb-webqsp}-\ref{responseucb-meta}. These results demonstrate that incorporating LLM-derived rewards accelerates convergence and improves performance compared with the setting without LLM-derived rewards. Among the tested ranges, the [0, 10] reward range achieves faster convergence and higher final performance than other ranges. Specifically, the performance of the method with LLM-derived rewards converges after approximately 3 iterations on both the WebQSP and MetaQA datasets, and after 5 iterations on the CWQ dataset. These findings indicate that the integration of LLM-derived rewards does not violate the boundedness or stationarity assumptions required by the convergence of UCB1. The number of iterations required to achieve converged performance can be set as a reliable computational budget for practical applications.

\subsubsection{Big-O Analysis} To assess the practical applicability of our method, we provide a Big-O analysis of our method and compare its complexity with both MCTS-based methods and other search strategies. 

Specifically, we follow the approach used in KBQA-o1 \cite{86} to assess the search space complexity of our method. Let $k$ denote the branching factor and $d$ the search depth. Our method operates through four stages: Selection, Expansion, Simulation, and Backpropagation. \textbf{Selection Stage:} The algorithm traverses the search tree up to depth \(d\), selecting the best node using UCB1 \cite{31}, resulting in a complexity of \(O(k \cdot d)\) per rollout. \textbf{Expansion Stage:} The \(B\) candidates are generated from \(k\) branches, introducing an additional complexity of \(O(\omega k)\), where \(\omega\) denotes the average search space complexity required for LLM evaluation and reflection, as well as similarity-matching operations, on each branch. \textbf{Simulation Stage:} This stage explores paths up to depth \(d\), contributing \(O(k \cdot d)\) complexity. \textbf{Backpropagation Stage:} The rewards are updated along the path, requiring \(O(d)\). The total search space complexity for a single rollout is \(O(k \cdot d)\). 
Therefore, with \(N\) rollouts, the overall search space complexity of our method is \(O(N \cdot k \cdot d)\), which scales linearly with both the number of rollouts and the search depth. Compared with existing MCTS-based KBQA methods such as KBQA-o1 \cite{86} and MCTS-KBQA \cite{88}, our method has the same search space. 
In contrast to other search strategies like DFS \cite{97} and BFS \cite{27}, which have exponential search complexity \(O(k^d)\), our method focuses on promising nodes, reducing unnecessary computation and maintaining efficiency.

\section{Conclusion}

This paper proposes the M-ER framework to address multi-hop KBQA challenges through synergistic integration of MCTS and LLM. Our approach overcomes the limitations of conventional search strategies by establishing a dynamic reasoning loop that combines semantic guidance with the selection of relevant nodes for the question. Experimental results on the WebQSP dataset demonstrate the M-ER's effectiveness in suppressing noisy relations while maintaining 78.88\% Hits@1.

\bibliographystyle{IEEEtran} 
\bibliography{tkde} 

@article{1,
  title={An introduction to question answering over linked data},
  author={Unger, Christina and Freitas, Andr{\'e} and Cimiano, Philipp},
  journal={Reasoning Web. Reasoning on the Web in the Big Data Era: 10th International Summer School 2014, Athens, Greece, September 8-13, 2014. Proceedings 10},
  pages={100--140},
  year={2014},
  publisher={Springer}
}

@inproceedings{2,
  title={Freebase: a collaboratively created graph database for structuring human knowledge},
  author={Bollacker, Kurt and Evans, Colin and Paritosh, Praveen and Sturge, Tim and Taylor, Jamie},
  booktitle={Proceedings of the 2008 ACM SIGMOD international conference on Management of data},
  pages={1247--1250},
  year={2008}
}

@article{3,
  title={Dbpedia--a large-scale, multilingual knowledge base extracted from wikipedia},
  author={Lehmann, Jens and Isele, Robert and Jakob, Max and Jentzsch, Anja and Kontokostas, Dimitris and Mendes, Pablo N and Hellmann, Sebastian and Morsey, Mohamed and Van Kleef, Patrick and Auer, S{\"o}ren and others},
  journal={Semantic web},
  volume={6},
  number={2},
  pages={167--195},
  year={2015},
  publisher={IOS Press}
}

@article{4,
  title={Knowledge base question answering with topic units},
  author={Lan, Yunshi and Wang, Shuohang and Jiang, Jing},
  year={2019}
}

@inproceedings{5,
  title={PullNet: Open Domain Question Answering with Iterative Retrieval on Knowledge Bases and Text},
  author={Sun, Haitian and Tania, Bedrax-Weiss and William, Cohen},
  booktitle={Proceedings of the 2019 Conference on Empirical Methods in Natural Language Processing and the 9th International Joint Conference on Natural Language Processing (EMNLP-IJCNLP)},
  pages={2380-2390},
  year={2019}
}

@inproceedings{6,
  title={Improving multi-hop knowledge base question answering by learning intermediate supervision signals},
  author={He, Gaole and Lan, Yunshi and Jiang, Jing and Zhao, Wayne Xin and Wen, Ji-Rong},
  booktitle={Proceedings of the 14th ACM international conference on web search and data mining},
  pages={553--561},
  year={2021}
}

@inproceedings{8,
  title={Query graph generation for answering multi-hop complex questions from knowledge bases},
  author={Lan, Yunshi and Jiang, Jing},
  year={2020},
  organization={Association for Computational Linguistics}
}

@article{9,
  title={Case-based reasoning for natural language queries over knowledge bases},
  author={Das, Rajarshi and Zaheer, Manzil and Thai, Dung and Godbole, Ameya and Perez, Ethan and Lee, Jay-Yoon and Tan, Lizhen and Polymenakos, Lazaros and McCallum, Andrew},
  journal={arXiv preprint arXiv:2104.08762},
  year={2021}
}

@inproceedings{10,
  title={Improving Complex Knowledge Base Question Answering with Relation-Aware Subgraph Retrieval and Reasoning Network},
  author={Luo, Dan and Sheng, Jiawei and Xu, Hongbo and Wang, Lihong and Wang, Bin},
  booktitle={2023 International Joint Conference on Neural Networks (IJCNN)},
  pages={1-8},
  year={2023},
  organization={IEEE}
}

@article{14,
  title={Enhancing Reasoning through Process Supervision with Monte Carlo Tree Search},
  author={Li, Shuangtao and Dong, Shuaihao and Luan, Kexin and Di, Xinhan and Ding, Chaofan},
  journal={arXiv preprint arXiv:2501.01478},
  year={2025}
}

@article{15,
  title={Airrag: Activating intrinsic reasoning for retrieval augmented generation via tree-based search},
  author={Feng, Wenfeng and Hao, Chuzhan and Zhang, Yuewei and Song, Jingyi and Wang, Hao},
  journal={arXiv preprint arXiv:2501.10053},
  year={2025}
}

@article{16,
  title={Technical report: Enhancing llm reasoning with reward-guided tree search},
  author={Jiang, Jinhao and Chen, Zhipeng and Min, Yingqian and Chen, Jie and Cheng, Xiaoxue and Wang, Jiapeng and Tang, Yiru and Sun, Haoxiang and Deng, Jia and Zhao, Wayne Xin and others},
  journal={arXiv preprint arXiv:2411.11694},
  year={2024}
}

@article{17,
  title={RARE: Retrieval-Augmented Reasoning Enhancement for Large Language Models},
  author={Tran, Hieu and Yao, Zonghai and Wang, Junda and Zhang, Yifan and Yang, Zhichao and Yu, Hong},
  journal={arXiv preprint arXiv:2412.02830},
  year={2024}
}

@inproceedings{18,
  title     = {SRA-MCTS: Self-driven Reasoning Augmentation with Monte Carlo Tree Search for Code Generation},
  author    = {Xu, Bin and Lin, Yiguan and Li, Yinghao and Gao, Yang},
  booktitle = {Proceedings of the Thirty-Fourth International Joint Conference on
               Artificial Intelligence, {IJCAI-25}},
  publisher = {International Joint Conferences on Artificial Intelligence Organization},
  editor    = {James Kwok},
  pages     = {8678--8686},
  year      = {2025},
  month     = {8},
  note      = {Main Track},
  doi       = {10.24963/ijcai.2025/965},
  url       = {https://doi.org/10.24963/ijcai.2025/965},
}

@article{22,
  title={Large language models as commonsense knowledge for large-scale task planning},
  author={Zhao, Zirui and Lee, Wee Sun and Hsu, David},
  journal={Advances in Neural Information Processing Systems},
  volume={36},
  pages={31967--31987},
  year={2023}
}

@article{23,
  title={Mutual reasoning makes smaller llms stronger problem-solvers},
  author={Qi, Zhenting and Ma, Mingyuan and Xu, Jiahang and Zhang, Li Lyna and Yang, Fan and Yang, Mao},
  journal={arXiv preprint arXiv:2408.06195},
  year={2024}
}

@inproceedings{24,
    title = "Open Domain Question Answering Using Early Fusion of Knowledge Bases and Text",
    author = "Sun, Haitian  and
      Dhingra, Bhuwan  and
      Zaheer, Manzil  and
      Mazaitis, Kathryn  and
      Salakhutdinov, Ruslan  and
      Cohen, William",
    editor = "Riloff, Ellen  and
      Chiang, David  and
      Hockenmaier, Julia  and
      Tsujii, Jun{'}ichi",
    booktitle = "Proceedings of the 2018 Conference on Empirical Methods in Natural Language Processing",
    month = oct # "-" # nov,
    year = "2018",
    address = "Brussels, Belgium",
    publisher = "Association for Computational Linguistics",
    url = "https://aclanthology.org/D18-1455/",
    doi = "10.18653/v1/D18-1455",
    pages = "4231--4242"
}

@inproceedings{27,
  title={ReasoningLM: Enabling Structural Subgraph Reasoning in Pre-trained Language Models for Question Answering over Knowledge Graph},
  author={Jiang, Jinhao and Zhou, Kun and Zhao, Wayne Xin and Li, Yaliang and Wen, Ji-Rong},
  booktitle={Proceedings of the 2023 Conference on Empirical Methods in Natural Language Processing},
  pages={3721--3735},
  year={2023}
}

@inproceedings{30,
  title={Integrating subgraph-aware relation and direction reasoning for question answering},
  author={Wang, Xu and Zhao, Shuai and Cheng, Bo and Han, Jiale and Yingting, Li and Yang, Hao and Sekulic, Ivan and Nan, Guoshun},
  booktitle={ICASSP 2021-2021 IEEE International Conference on Acoustics, Speech and Signal Processing (ICASSP)},
  pages={7808--7812},
  year={2021},
  organization={IEEE}
}

@article{31,
  title={Finite-time analysis of the multiarmed bandit problem},
  author={Auer, Peter and Cesa-Bianchi, Nicolo and Fischer, Paul},
  journal={Machine learning},
  volume={47},
  pages={235--256},
  year={2002},
  publisher={Springer}
}

@article{32,
  title={Rest-mcts*: Llm self-training via process reward guided tree search},
  author={Zhang, Dan and Zhoubian, Sining and Hu, Ziniu and Yue, Yisong and Dong, Yuxiao and Tang, Jie},
  journal={Advances in Neural Information Processing Systems},
  volume={37},
  pages={64735--64772},
  year={2024}
}

@inproceedings{35,
  title={The value of semantic parse labeling for knowledge base question answering},
  author={Yih, Wen-tau and Richardson, Matthew and Meek, Christopher and Chang, Ming-Wei and Suh, Jina},
  booktitle={Proceedings of the 54th Annual Meeting of the Association for Computational Linguistics (Volume 2: Short Papers)},
  pages={201--206},
  year={2016}
}

@article{36,
  title={Event Assignment Based on KBQA for Government Service Hotlines},
  author={Wang, Yongzhang and Lian, Defu and Chen, Enhong},
  journal={Applied Artificial Intelligence},
  volume={38},
  number={1},
  pages={2348162},
  year={2024},
  publisher={Taylor \& Francis}
}

@article{37,
  title={The web as a knowledge-base for answering complex questions},
  author={Talmor, Alon and Berant, Jonathan},
  journal={arXiv preprint arXiv:1803.06643},
  year={2018}
}

@article{39,
  title={Lighter And Better: Towards Flexible Context Adaptation For Retrieval Augmented Generation},
  author={Liu, Zheng and Wu, Chenyuan and Shao, Ninglu and Xiao, Shitao and Li, Chaozhuo and Lian, Defu},
  journal={arXiv preprint arXiv:2409.15699},
  year={2024}
}

@inproceedings{40,
  title={Key-Value Memory Networks for Directly Reading Documents},
  author={Miller, Alexander and Fisch, Adam and Dodge, Jesse and Karimi, Amir-Hossein and Bordes, Antoine and Weston, Jason},
  booktitle={Proceedings of the 2016 Conference on Empirical Methods in Natural Language Processing},
  year={2016},
  organization={Association for Computational Linguistics}
}

@inproceedings{41,
  title={Improving multi-hop question answering over knowledge graphs using knowledge base embeddings},
  author={Saxena, Apoorv and Tripathi, Aditay and Talukdar, Partha},
  booktitle={Proceedings of the 58th annual meeting of the association for computational linguistics},
  pages={4498--4507},
  year={2020}
}

@article{44,
  title={Subgraph retrieval enhanced model for multi-hop knowledge base question answering},
  author={Zhang, Jing and Zhang, Xiaokang and Yu, Jifan and Tang, Jian and Tang, Jie and Li, Cuiping and Chen, Hong},
  journal={arXiv preprint arXiv:2202.13296},
  year={2022}
}

@inproceedings{45,
  title={SPARQA: skeleton-based semantic parsing for complex questions over knowledge bases},
  author={Sun, Yawei and Zhang, Lingling and Cheng, Gong and Qu, Yuzhong},
  booktitle={Proceedings of the AAAI conference on artificial intelligence},
  volume={34},
  number={05},
  pages={8952--8959},
  year={2020}
}

@inproceedings{46,
  title={Answering complex questions by combining information from curated and extracted knowledge bases},
  author={Bhutani, Nikita and Zheng, Xinyi and Qian, Kun and Li, Yunyao and Jagadish, H},
  booktitle={Proceedings of the first workshop on natural language interfaces},
  pages={1--10},
  year={2020}
}

@inproceedings{47,
  title={Structure-Aware Reasoning for Knowledge Base Question Answering},
  author={Ma, Lu and Zhang, Peng and Zhu, Xi and Luo, Dan and Wang, Bin},
  booktitle={Pacific-Asia Conference on Knowledge Discovery and Data Mining},
  pages={562--573},
  year={2022},
  organization={Springer}
}

@article{48,
  title={Improving embedded knowledge graph multi-hop question answering by introducing relational chain reasoning},
  author={Jin, Weiqiang and Zhao, Biao and Yu, Hang and Tao, Xi and Yin, Ruiping and Liu, Guizhong},
  journal={Data Mining and Knowledge Discovery},
  volume={37},
  number={1},
  pages={255--288},
  year={2023},
  publisher={Springer}
}

@article{50,
  title={Think-on-graph: Deep and responsible reasoning of large language model on knowledge graph},
  author={Sun, Jiashuo and Xu, Chengjin and Tang, Lumingyuan and Wang, Saizhuo and Lin, Chen and Gong, Yeyun and Ni, Lionel M and Shum, Heung-Yeung and Guo, Jian},
  journal={arXiv preprint arXiv:2307.07697},
  year={2023}
}

@inproceedings{49,
  title={Generate-on-Graph: Treat LLM as both Agent and KG for Incomplete Knowledge Graph Question Answering},
  author={Xu, Yao and He, Shizhu and Chen, Jiabei and Wang, Zihao and Song, Yangqiu and Tong, Hanghang and Liu, Guang and Zhao, Jun and Liu, Kang},
  booktitle={Proceedings of the 2024 Conference on Empirical Methods in Natural Language Processing},
  pages={18410--18430},
  year={2024}
}

@inproceedings{53,
    title = "{S}truct{GPT}: A General Framework for Large Language Model to Reason over Structured Data",
    author = "Jiang, Jinhao  and
      Zhou, Kun  and
      Dong, Zican  and
      Ye, Keming  and
      Zhao, Xin  and
      Wen, Ji-Rong",
    editor = "Bouamor, Houda  and
      Pino, Juan  and
      Bali, Kalika",
    booktitle = "Proceedings of the 2023 Conference on Empirical Methods in Natural Language Processing",
    month = dec,
    year = "2023",
    address = "Singapore",
    publisher = "Association for Computational Linguistics",
    url = "https://aclanthology.org/2023.emnlp-main.574/",
    doi = "10.18653/v1/2023.emnlp-main.574",
    pages = "9237--9251",
}

@article{54,
    author = {Chung, Hyung Won and Hou, Le and Longpre, Shayne and Zoph, Barret and Tai, Yi and Fedus, William and Li, Yunxuan and Wang, Xuezhi and Dehghani, Mostafa and Brahma, Siddhartha and Webson, Albert and Gu, Shixiang Shane and Dai, Zhuyun and Suzgun, Mirac and Chen, Xinyun and Chowdhery, Aakanksha and Castro-Ros, Alex and Pellat, Marie and Robinson, Kevin and Valter, Dasha and Narang, Sharan and Mishra, Gaurav and Yu, Adams and Zhao, Vincent and Huang, Yanping and Dai, Andrew and Yu, Hongkun and Petrov, Slav and Chi, Ed H. and Dean, Jeff and Devlin, Jacob and Roberts, Adam and Zhou, Denny and Le, Quoc V. and Wei, Jason},
    title = {Scaling instruction-finetuned language models},
    year = {2025},
    issue_date = {January 2024},
    publisher = {JMLR.org},
    volume = {25},
    number = {1},
    issn = {1532-4435},
    journal = {J. Mach. Learn. Res.},
    month = mar,
    articleno = {70},
    numpages = {53}
}

@inproceedings{59,
    title = "{FC}-{KBQA}: A Fine-to-Coarse Composition Framework for Knowledge Base Question Answering",
    author = "Zhang, Lingxi  and
      Zhang, Jing  and
      Wang, Yanling  and
      Cao, Shulin  and
      Huang, Xinmei  and
      Li, Cuiping  and
      Chen, Hong  and
      Li, Juanzi",
    editor = "Rogers, Anna  and
      Boyd-Graber, Jordan  and
      Okazaki, Naoaki",
    booktitle = "Proceedings of the 61st Annual Meeting of the Association for Computational Linguistics (Volume 1: Long Papers)",
    month = jul,
    year = "2023",
    address = "Toronto, Canada",
    publisher = "Association for Computational Linguistics",
    url = "https://aclanthology.org/2023.acl-long.57/",
    doi = "10.18653/v1/2023.acl-long.57",
    pages = "1002--1017"
}

@misc{60,
      title={Knowledge Base Question Answering by Case-based Reasoning over Subgraphs}, 
      author={Rajarshi Das and Ameya Godbole and Ankita Naik and Elliot Tower and Robin Jia and Manzil Zaheer and Hannaneh Hajishirzi and Andrew McCallum},
      year={2022},
      eprint={2202.10610},
      archivePrefix={arXiv},
      primaryClass={cs.CL},
      url={https://arxiv.org/abs/2202.10610}, 
}

@inproceedings{61,
author = {Bhutani, Nikita and Zheng, Xinyi and Jagadish, H V},
title = {Learning to Answer Complex Questions over Knowledge Bases with Query Composition},
year = {2019},
isbn = {9781450369763},
publisher = {Association for Computing Machinery},
address = {New York, NY, USA},
url = {https://doi.org/10.1145/3357384.3358033},
doi = {10.1145/3357384.3358033},
booktitle = {Proceedings of the 28th ACM International Conference on Information and Knowledge Management},
pages = {739–748},
numpages = {10},
location = {Beijing, China},
series = {CIKM '19}
}

@inproceedings{63,
  title={Generate-then-Ground in Retrieval-Augmented Generation for Multi-hop Question Answering},
  author={Shi, Zhengliang and Zhang, Shuo and Sun, Weiwei and Gao, Shen and Ren, Pengjie and Chen, Zhumin and Ren, Zhaochun},
  booktitle={Proceedings of the 62nd Annual Meeting of the Association for Computational Linguistics (Volume 1: Long Papers)},
  pages={7339--7353},
  year={2024}
}

@inproceedings{64,
  title={ChatKBQA: A Generate-then-Retrieve Framework for Knowledge Base Question Answering with Fine-tuned Large Language Models},
  author={Luo, Haoran and Haihong, E and Tang, Zichen and Peng, Shiyao and Guo, Yikai and Zhang, Wentai and Ma, Chenghao and Dong, Guanting and Song, Meina and Lin, Wei and others},
  booktitle={Findings of the Association for Computational Linguistics ACL 2024},
  pages={2039--2056},
  year={2024}
}

@inproceedings{65,
  title={Few-shot Transfer Learning for Knowledge Base Question Answering: Fusing Supervised Models with In-Context Learning},
  author={Patidar, Mayur and Sawhney, Riya and Singh, Avinash and Chatterjee, Biswajit and Bhattacharya, Indrajit and others},
  booktitle={Proceedings of the 62nd Annual Meeting of the Association for Computational Linguistics (Volume 1: Long Papers)},
  pages={9147--9165},
  year={2024}
}

@article{76,
  title={A Method of Query Graph Reranking for Knowledge Base Question Answering},
  author={Jia, Yonghui and Chen, Wenliang},
  journal={arXiv preprint arXiv:2204.12808},
  year={2022}
}

@inproceedings{77,
  title={A Two-Stage Approach towards Generalization in Knowledge Base Question Answering},
  author={Ravishankar, Srinivas and Thai, Dung and Abdelaziz, Ibrahim and Mihindukulasooriya, Nandana and Naseem, Tahira and Kapanipathi, Pavan and Rossiello, Gaetano and Fokoue-Nkoutche, Achille},
  booktitle={Findings of the Association for Computational Linguistics: EMNLP 2022},
  pages={5571--5580},
  year={2022}
}

@inproceedings{78,
  title={An in-context schema understanding method for knowledge base question answering},
  author={Liu, Yantao and Li, Zixuan and Jin, Xiaolong and Guo, Yucan and Bai, Long and Guan, Saiping and Guo, Jiafeng and Cheng, Xueqi},
  booktitle={International Conference on Knowledge Science, Engineering and Management},
  pages={419--434},
  year={2024},
  organization={Springer}
}

@article{81,
  title={Cost-efficient knowledge-based question answering with large language models},
  author={Dong, Junnan and Zhang, Qinggang and Zhou, Chuang and Chen, Hao and Zha, Daochen and Huang, Xiao},
  journal={arXiv preprint arXiv:2405.17337},
  year={2024}
}

@inproceedings{83,
  title={Flexkbqa: A flexible llm-powered framework for few-shot knowledge base question answering},
  author={Li, Zhenyu and Fan, Sunqi and Gu, Yu and Li, Xiuxing and Duan, Zhichao and Dong, Bowen and Liu, Ning and Wang, Jianyong},
  booktitle={Proceedings of the AAAI conference on artificial intelligence},
  volume={38},
  number={17},
  pages={18608--18616},
  year={2024}
}

@article{86,
  title={KBQA-o1: Agentic Knowledge Base Question Answering with Monte Carlo Tree Search},
  author={Luo, Haoran and Guo, Yikai and Lin, Qika and Wu, Xiaobao and Mu, Xinyu and Liu, Wenhao and Song, Meina and Zhu, Yifan and Tuan, Luu Anh and others},
  journal={arXiv preprint arXiv:2501.18922},
  year={2025}
}

@article{88,
  title={MCTS-KBQA: Monte Carlo Tree Search for Knowledge Base Question Answering},
  author={Xiong, Guanming and Li, Haochen and Zhao, Wen},
  journal={arXiv preprint arXiv:2502.13428},
  year={2025}
}

@inproceedings{89,
  title={React: Synergizing reasoning and acting in language models},
  author={Yao, Shunyu and Zhao, Jeffrey and Yu, Dian and Du, Nan and Shafran, Izhak and Narasimhan, Karthik and Cao, Yuan},
  booktitle={International Conference on Learning Representations (ICLR)},
  year={2023}
}

@article{90,
  title={Reasoning on graphs: Faithful and interpretable large language model reasoning},
  author={Luo, Linhao and Li, Yuan-Fang and Haffari, Gholamreza and Pan, Shirui},
  journal={arXiv preprint arXiv:2310.01061},
  year={2023}
}

@inproceedings{91,
  title={Interactive-KBQA: Multi-Turn Interactions for Knowledge Base Question Answering with Large Language Models},
  author={Xiong, Guanming and Bao, Junwei and Zhao, Wen},
  booktitle={Proceedings of the 62nd Annual Meeting of the Association for Computational Linguistics (Volume 1: Long Papers)},
  pages={10561--10582},
  year={2024}
}

@inproceedings{94,
  title={Rgr-kbqa: Generating logical forms for question answering using knowledge-graph-enhanced large language model},
  author={Feng, Tengfei and He, Liang},
  booktitle={Proceedings of the 31st International Conference on Computational Linguistics},
  pages={3057--3070},
  year={2025}
}

@inproceedings{95,
  title={GRV-KBQA: A Three-Stage Framework for Knowledge Base Question Answering with Decoupled Logical Structure, Semantic Grounding and Structure-Aware Validation},
  author={Tian, Yuhang and Yang, Pan and Song, Dandan and Wu, Zhijing and Wang, Hao},
  booktitle={Findings of the Association for Computational Linguistics: EMNLP 2025},
  pages={2618--2632},
  year={2025}
}

@inproceedings{96,
  title={Variational reasoning for question answering with knowledge graph},
  author={Zhang, Yuyu and Dai, Hanjun and Kozareva, Zornitsa and Smola, Alexander and Song, Le},
  booktitle={Proceedings of the AAAI conference on artificial intelligence},
  volume={32},
  number={1},
  year={2018}
}

@article{97,
  title={SmartDoc: A Context-Aware Agentic Method Comment Generation Plugin},
  author={Etemadi, Vahid and Robles, Gregorio},
  journal={arXiv preprint arXiv:2511.00450},
  year={2025}
}

@inproceedings{98,
  title={Dense passage retrieval for open-domain question answering},
  author={Karpukhin, Vladimir and O{\u{g}}uz, Barlas and Min, Sewon and Lewis, Patrick and Wu, Ledell and Edunov, Sergey and Chen, Danqi and Yih, Wen Tau},
  booktitle={2020 Conference on Empirical Methods in Natural Language Processing, EMNLP 2020},
  pages={6769--6781},
  year={2020},
  organization={Association for Computational Linguistics (ACL)}
}

@inproceedings{99,
  title={C-pack: Packed resources for general chinese embeddings},
  author={Xiao, Shitao and Liu, Zheng and Zhang, Peitian and Muennighoff, Niklas and Lian, Defu and Nie, Jian-Yun},
  booktitle={Proceedings of the 47th international ACM SIGIR conference on research and development in information retrieval},
  pages={641--649},
  year={2024}
}

@article{102,
  title={RPM-MCTS: Knowledge-Retrieval as Process Reward Model with Monte Carlo Tree Search for Code Generation},
  author={Lin, Yuanyuan and Ouyang, Xiangyu and Zhang, Teng and Sui, Kaixin},
  journal={arXiv preprint arXiv:2511.19895},
  year={2025}
}

@article{103,
  title={Magellan: Guided MCTS for Latent Space Exploration and Novelty Generation},
  author={Chang, Lufan},
  journal={arXiv preprint arXiv:2510.21341},
  year={2025}
}

@article{104,
  title={Empowering RepoQA-Agent based on Reinforcement Learning Driven by Monte-carlo Tree Search},
  author={Li, Guochang and Liu, Yuchen and Qin, Zhen and Wang, Yunkun and Zhong, Jianping and Zhi, Chen and Li, Binhua and Huang, Fei and Li, Yongbin and Deng, Shuiguang},
  journal={arXiv preprint arXiv:2510.26287},
  year={2025}
}

@misc{105,
      title={DeepSeek-V3 Technical Report}, 
      author={DeepSeek-AI},
      year={2024},
      eprint={2412.19437},
      archivePrefix={arXiv},
      primaryClass={cs.CL},
      url={https://arxiv.org/abs/2412.19437}, 
}

@article{106,
  title={Challenges and Applications of Large Language Models: A Comparison of GPT and DeepSeek family of models},
  author={Sharma, Shubham and Tuli, Sneha and Badam, Narendra},
  journal={arXiv preprint arXiv:2508.21377},
  year={2025}
}

@ARTICLE{107,
  author={Chen, Xiaoshu and Zhou, Sihang and Liang, Ke and Liu, Xinwang},
  journal={IEEE Transactions on Neural Networks and Learning Systems}, 
  title={Distilling Reasoning Ability From Large Language Models With Adaptive Thinking}, 
  year={2025},
  volume={36},
  number={11},
  pages={19820-19833},
  doi={10.1109/TNNLS.2025.3591266}}

@article{108,
  title={Putting on the Thinking Hats: A Survey on Chain of Thought Fine-tuning from the Perspective of Human Reasoning Mechanism},
  author={Chen, Xiaoshu and Zhou, Sihang and Liang, Ke and Yuan, Duanyang and Chen, Haoyuan and Sun, Xiaoyu and Meng, Linyuan and Liu, Xinwang},
  journal={arXiv preprint arXiv:2510.13170},
  year={2025}
}

@article{109,
  title={From concrete to abstract: multi-view clustering on relational knowledge},
  author={Liang, Ke and Meng, Lingyuan and Li, Hao and Wang, Jun and Lan, Long and Li, Miaomiao and Liu, Xinwang and Wang, Huaimin},
  journal={IEEE Transactions on Pattern Analysis and Machine Intelligence},
  year={2025},
  publisher={IEEE}
}

@inproceedings{110,
  title={Learn from relational correlations and periodic events for temporal knowledge graph reasoning},
  author={Liang, Ke and Meng, Lingyuan and Liu, Meng and Liu, Yue and Tu, Wenxuan and Wang, Siwei and Zhou, Sihang and Liu, Xinwang},
  booktitle={Proceedings of the 46th international ACM SIGIR conference on research and development in information retrieval},
  pages={1559--1568},
  year={2023}
}

\begin{IEEEbiography}
[{\includegraphics[width=0.75in,height=0.9in,clip,keepaspectratio]{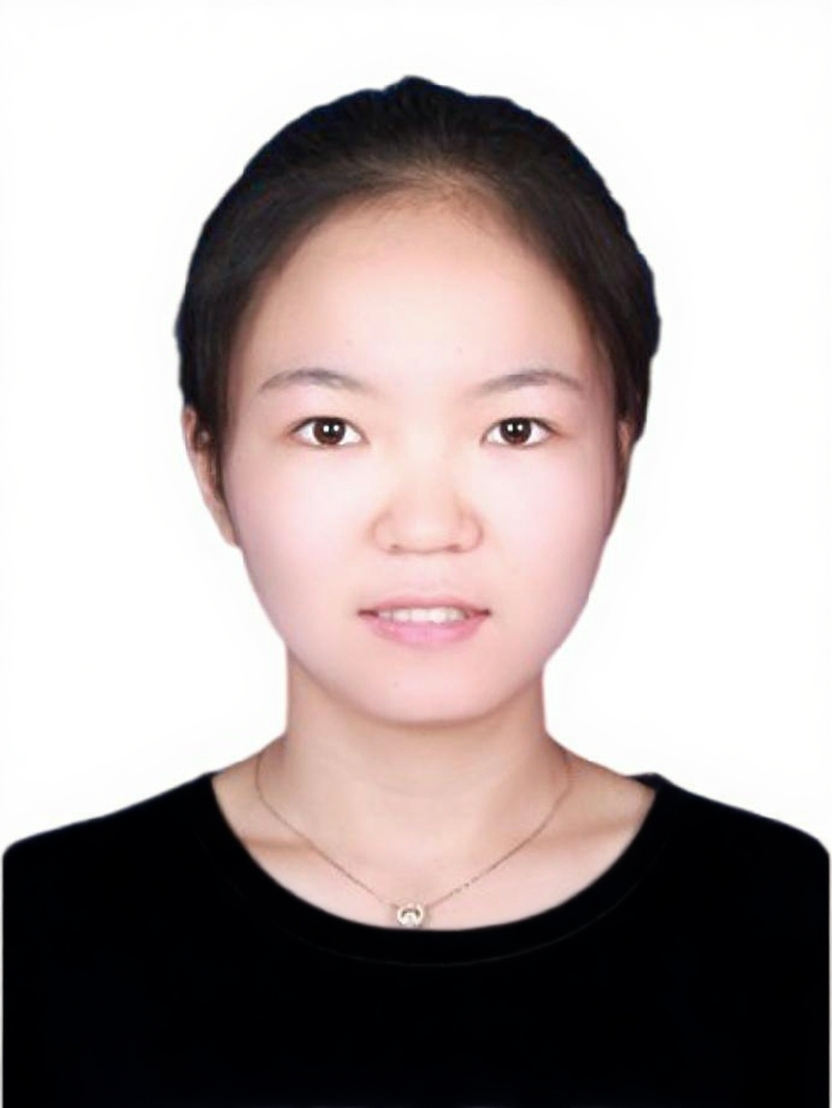}}]{Duanyang Yuan} is currently pursuing a Ph.D. degree at the National University of Defense Technology (NUDT).  Her research interests include knowledge graphs, large language models, information retrieval, and retrieval-augmented generation. She has published several papers in AAAI, Frontiers in Physiology, etc.
\end{IEEEbiography}

\begin{IEEEbiography}
[{\includegraphics[width=0.75in,height=0.9in,clip,keepaspectratio]{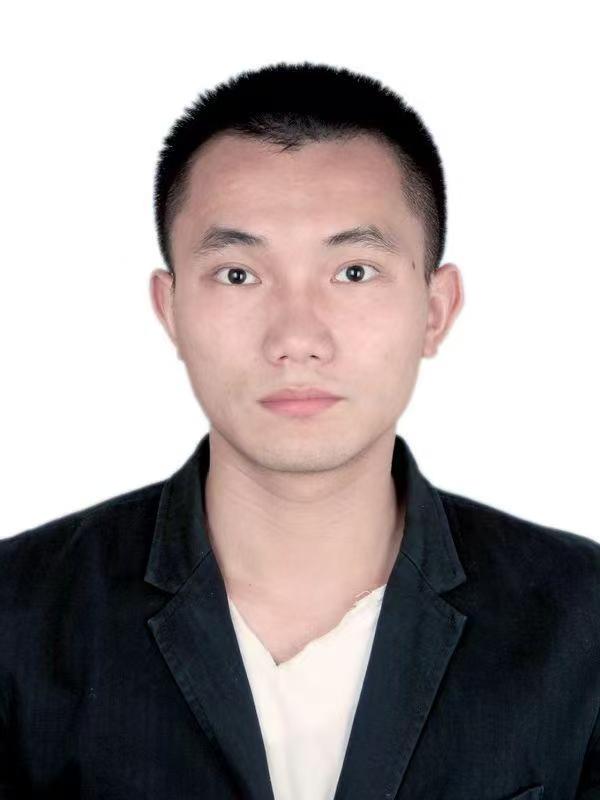}}]{Sihang Zhou} received his PhD degree from School of Computer, National University of Defense Technology (NUDT), China. He is now an associate professor at the College of Intelligence Science and Technology, NUDT. His current research interests include machine learning, knowledge graph, and medical image analysis. Dr. Zhou has published 40+ peer-reviewed papers, including IEEE T-IP, IEEE T-NNLS, IEEE T-MI, Medical Image Analysis, AAAI, MICCAI, etc.
\end{IEEEbiography}

\begin{IEEEbiography}[{\includegraphics[width=0.75in,height=0.9in,clip,keepaspectratio]{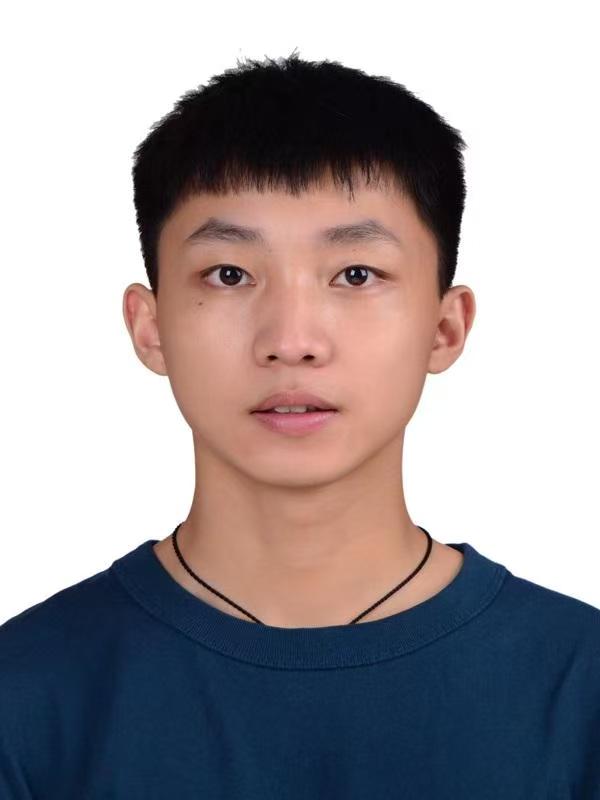}}] {Xiaoshu Chen} is currently pursuing a Ph.D. degree at the National University of Defense Technology (NUDT). Before joining NUDT, he got his MSc degree at WuHan University (WHU) and BSc degree at China University of Mining and Technology (CUMT). His research interests include large language models, knowledge graph, information retrieval and semantic segmentation. He has published several papers in AAAI, IEEE TGRS and Neurocomputing, etc.
\end{IEEEbiography}

\begin{IEEEbiography}[{\includegraphics[width=0.75in,height=0.9in,clip,keepaspectratio]{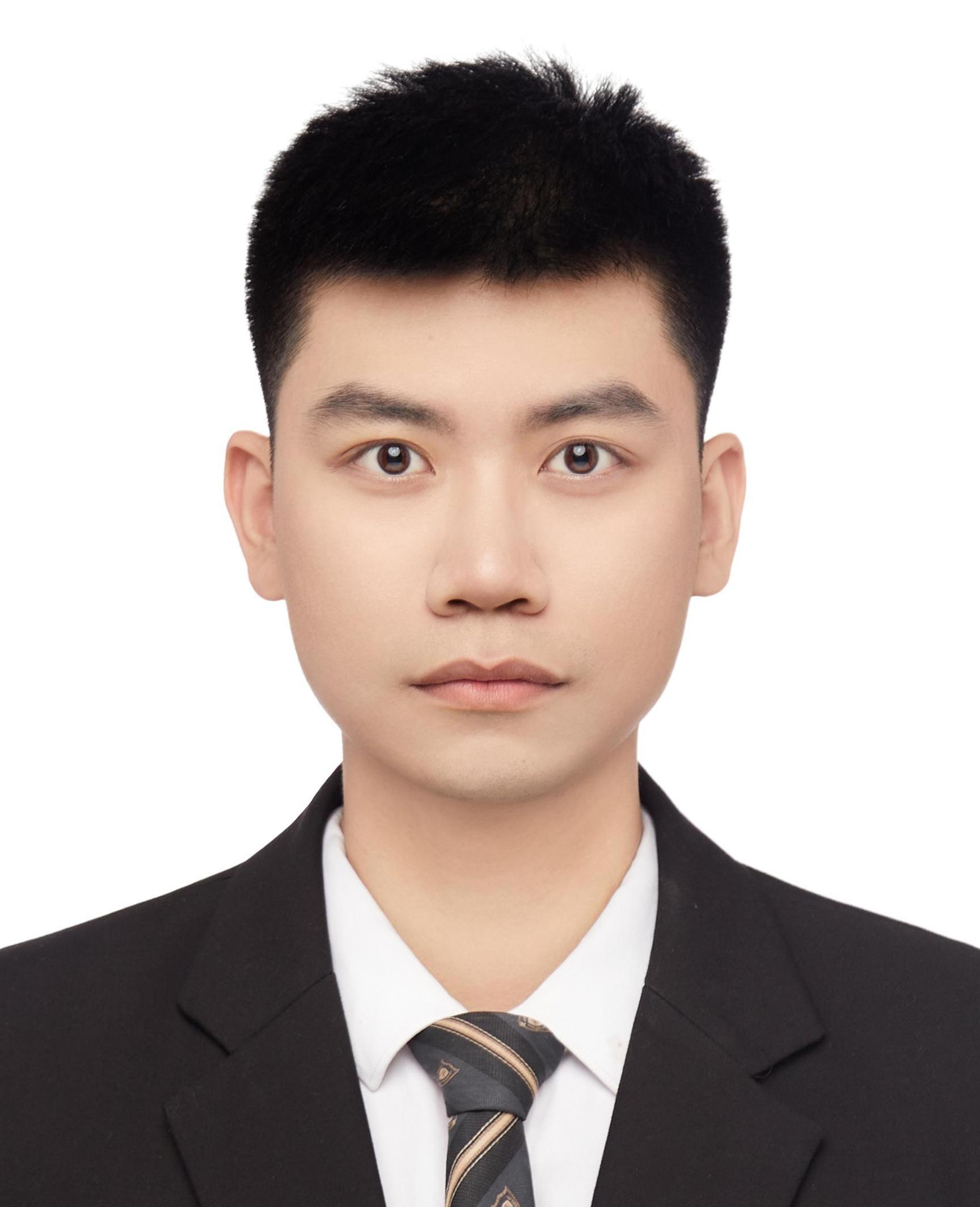}}]
{Haoyuan Chen} is currently pursuing a Ph.D. degree at the National University of Defense Technology (NUDT). His research interests include large language models, retrieval-augmented generation, and multi-agent systems.
\end{IEEEbiography}

\begin{IEEEbiography}[{\includegraphics[width=0.75in,height=0.9in,clip,keepaspectratio]{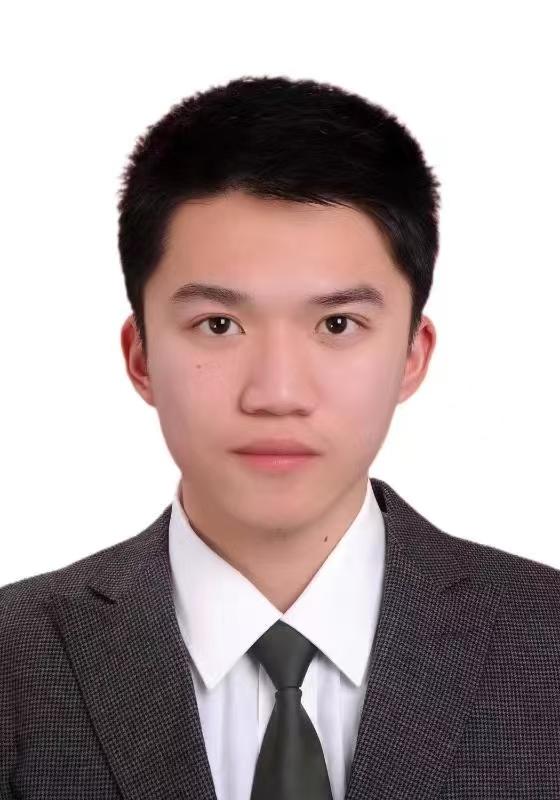}}]
{Ke Liang} received his Ph.D. degree at the National University of Defense Technology (NUDT). He got his BSc degree at Beihang University (BUAA) and received his MSc degree from the Pennsylvania State University (PSU). His current research interests include knowledge graphs, graph learning, and healthcare AI. He has published several papers in highly regarded journals and conferences such as SIGIR, AAAI, ICML, ACM MM, IEEE TNNLS, IEEE TKDE, etc.
\end{IEEEbiography}

\begin{IEEEbiography}
[{\includegraphics[width=0.75in,height=0.9in,clip,keepaspectratio]{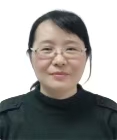}}] 
{Jian Huang} received the Ph.D. degree from the College of Mechatronics and Automation, National University of Defense Technology, Changsha, China, in 2000. She is now a professor at the College of Intelligence Science, National University of Defense Technology, China. Her current research interests include system simulation and artificial intelligence.
\end{IEEEbiography}

\vfill

\end{document}